\documentclass[12pt]{article}

\pdfoutput=1 
\newcommand{\blind}{0}

\usepackage{amsmath,amsfonts,amssymb}
\usepackage{latexsym,textcomp,cancel}
\usepackage{graphicx}
\usepackage{natbib}
\usepackage{appendix}
\usepackage{url,color}
\usepackage{multirow, rotating}
\usepackage{cancel}
\usepackage{enumerate}

\newcommand{\bftheta} {\boldsymbol{\theta}}

\newcommand{\bfTheta} {\boldsymbol{\Theta}}

\newcommand{\bfv} {\mathbf{v}}
\newcommand{\bfb} {\mathbf{b}}

\newcommand{\bfX} {\mathbf{X}}
\newcommand{\bfY} {\mathbf{Y}}

\newcommand{\bfL}{\mathbf{L}}
\newcommand{\bfI}{\mathbf{I}}

\renewcommand{\Pr}{\mathsf{Pr}}

\newcommand{\V}{\mathsf{Var}}

\DeclareMathOperator{\sgn}{sgn}

\newcommand{\normal}{\mathsf{Normal}}

\newcommand{\DE}{\mathsf{DE}}

\newcommand{\Gam}{\mathsf{Gam}}

\newcommand{\IG}{\mathsf{IGau}}

\newcommand{\PG}{\mathsf{PG}}
\newcommand{\TN}{\mathsf{TN}}

\begin{document}

\def\spacingset#1{\renewcommand{\baselinestretch}%
{#1}\small\normalsize} \spacingset{1}

\if0\blind
{
  \title{\bf Bayesian Fused Lasso regression for dynamic binary networks}
  \author{Brenda Betancourt
  \hspace{.2cm}\\
    Department of Statistics, UC Santa Cruz\\
     and \\
    Abel Rodr{\'i}guez  \\
    Department of Statistics, UC Santa Cruz\\
    and \\
    Naomi Boyd  \\
    Department of Finance, West Virginia University}
  \maketitle
} \fi

\if1\blind
{
  \bigskip
  \bigskip
  \bigskip
  \begin{center}
    {\LARGE\bf Bayesian Fused Lasso regression for dynamic binary networks}
\end{center}
  \medskip
} \fi

\bigskip
\begin{abstract}
We propose a multinomial logistic regression model for link prediction in a time series of directed binary networks. To account for the dynamic nature of the data we employ a dynamic model for the model parameters that is strongly connected with the fused lasso penalty. In addition to promoting sparseness, this prior allows us to explore the presence of change points in the structure of the network. We introduce fast computational algorithms for estimation and prediction using both optimization and Bayesian approaches. The performance of the model is illustrated using simulated data and data from a financial trading network in the NYMEX natural gas futures market. Supplementary material containing the trading network data set and code to implement the algorithms is available online.
\end{abstract}

\noindent%
{\it Keywords:} multinomial logistic regression, network link prediction, P\'{o}lya-Gamma latent variables,
Split Bregman method.
\vfill

\newpage
\spacingset{1} 

\section{Introduction}\label{sec:intro}


Network data, in which observations correspond to the interactions among a group of nodes, has become pervasive in disciplines as diverse as social, physical and biological sciences. Accordingly, there has been a growing interest in developing tools for the analysis of network data, particularly from a model-based perspective (for excellent reviews see \citealp{Newman}, \citealp{GoldZhAi09} and \citealp{Snij11}). The focus on this paper is on models for time series of binary directed networks that involve the same set of subjects at each time point. In particular, our work is motivated by the study of financial trading networks (FTNs), which capture the pattern of buy and sell transactions in a financial market. A primary goal in the analysis of this type of dynamic network data is link prediction at future times, going as far as predicting the structure of the whole network. An additional goal is to provide a simple model to explore the evolution of the network, and possibly identify change-points in the network dynamic. To accomplish these goals we extend the idea of $p_1$ models initially proposed by \citet{Holland} for static binary networks.

Consider a directed binary network among $n$ nodes, $\bfY= [ y_{i,j} ]$, where $y_{i,j} = 1$ if there is a link directed from node $i$ to node $j$, and $y_{i,j} = 0$ otherwise. Holland and Leinhardt's model assumes conditional independence between pairs of nodes (dyads) and focuses on modeling the pairs $( y_{i,j},y_{j,i} )$ jointly for $i<j$, $j=1,\ldots,n$ as follows
\begin{align}\label{eq:p1model}
p\left( y_{i,j}, y_{j,i} \right) \propto 
\exp  \left\{ \theta_{1} y_{i,j} + \theta_{2} y_{j,i}+\theta_{3} y_{i,j}y_{j,i} \right\}.
\end{align}

This class of models has been extended to a dynamic setting by introducing Markov dependency upon past observations (e.g., see \citealp{BaCa96}). In contrast, in the modeling approach discussed in this paper, the model parameters are set to be time dependent to add flexibility and account for alterations in the network evolution over time. One challenging feature that is often present in model-based approaches to network data is high-dimensionality. In particular, the number of parameters in our proposed model is larger than the number of available observations. To deal with this issue we resort to fused lasso regression by imposing an $L_1$ penalty on the difference between neighboring model parameters \citep{Tibshi05}. In a Bayesian setting, this is equivalent to assuming a double exponential prior on the differences of the coefficients in contiguous time points. Here, we explore two different computational approaches for our model. First, full Bayesian inference is presented and implemented using two different sampling schemes. However, the heavy computational load of a full Bayesian analysis is a challenging task as the number of nodes in the network increases. As an alternative, we also carry out maximum a posteriori (MAP) estimation utilizing an optimization approach. 

The remainder of the paper is organized as follows: Section \ref{sec:model} describes our modeling approach. Section \ref{sec:estimation} describes the computational algorithms for estimation and prediction from optimization and Bayesian perspectives. Section \ref{sec:related} discusses other related work. Section \ref{sec:applications} presents three illustrations, two based on simulated data and a third one that focuses on trading networks from the natural gas futures market in the New York Mercantile Exchange (NYMEX). Finally, a short discussion is presented in Section \ref{sec:discussion}.

\section{Modeling Approach}\label{sec:model}

Consider a sequence of binary directed networks $\bfY_{1},\ldots,\bfY_{T}$, each one observed over a common set of $n$ nodes. The adjacency matrix of the network at time $t$ is therefore an $n \times n$ binary matrix $\bfY_{t} = [ y_{i,j,t} ]$, where $y_{i,j,t} = 1$ if there is a link directed from node $i$ to node $j$ at time $t$, and $y_{i,j,t} = 0$ otherwise. We adopt the convention $y_{i,i,t} \equiv 0$ so that there are no loops within the network. In the illustration we discuss in Section \ref{sec:NYMEX}, the nodes in the network correspond to traders in the NYMEX natural gas futures market, so that $y_{i,j,t} = 1$ if trader $i$ sold to trader $j$ at least once during week $t$.

We consider an extension of \eqref{eq:p1model} in which the pairs $\{ ( y_{i,j,t} , y_{j,i,t} ) : i < j \}$ are modeled  independently using a logistic model of the form
\begin{equation}\label{eq:model}
p\left( y_{i,j,t},y_{j,i,t} \right) \propto
\exp  \left\{\theta_{i,j,t,1} y_{i,j,t} +  \theta_{i,j,t,2} y_{j,i,t} + \theta_{i,j,t,3} y_{i,j,t}y_{j,i,t} \right\} ,
\end{equation}
where $\theta_{i,j,t,1}$ and $\theta_{i,j,t,2}$ represent the baseline probabilities of a directed link between nodes $i$ and $j$, and $\theta_{i,j,t,3}$ controls the level of dependence between $y_{i,j,t}$ and $y_{j,i,t}$. For example, $\theta_{i,j,t,3} = 0$ implies that $y_{i,j,t}$ and $y_{j,i,t}$ are conditionally independent with $\Pr( y_{i,j,t} = 1 ) = \exp\left\{ \theta_{i,j,t,1} \right\}/\left(1 + \exp\left\{ \theta_{i,j,t,1} \right\} \right)$ and $\Pr( y_{j,i,t} = 1 ) = \exp\left\{ \theta_{i,j,t,2} \right\}/\left(1 + \exp\left\{ \theta_{i,j,t,2} \right\} \right)$. On the other hand, $\theta_{i,j,t,3} > 0$ favors outcomes in which $y_{i,j,t} = y_{j,i,t}$ (a phenomenon often called positive reciprocity in the network literature), while $\theta_{i,j,t,3} < 0$ favors situations in which $y_{i,j,t} \ne y_{j,i,t}$ (often called negative reciprocity).  Hence, by allowing the values of $y_{i,j,t}$ and $y_{j,i,t}$ to be potentially correlated the model can accommodate reciprocity.  

The parameters in the multinomial logistic model we just described are time dependent.  Hence, it is natural and useful to take into account the information about their temporal correlation structure in the estimation process.  In particular, we are interested in a random walk model with double exponential priors of the form:
\begin{align*}
\theta_{i,j,t,r}& = \theta_{i,j,t-1,r}+\epsilon_{i,j,t,r}, & \epsilon_{i,j,t,r} &\sim \DE( 0,1/\lambda ),
\end{align*}
where $\DE$ represents the double exponential distribution, and  $\lambda> 0$ is the parameter that controls the shrinkage level in the differences of the coefficients.  A dynamic model of this type on the parameters leads to the joint prior
\begin{align*}
p \left( \bfTheta_{i,j,r} \mid \lambda \right) \propto \exp\left\{ -\lambda\sum_{t=1}^{T}|\theta_{i,j,t,r}-\theta_{i,j,t-1,r}|\right\},
\end{align*}
where $\bfTheta_{i,j,r}=\left(\theta_{i,j,0,r}, \theta_{i,j,1,r},\ldots, \theta_{i,j,T,r}\right)$ is the vector of parameters for class $r$ and pair of nodes $(i,j)$, and $\theta_{i,j,0,r}=\hat{\theta}_{r,0}$ is assumed known.  This pairwise difference prior belongs to the class of Markov random fields and corresponds to a scale mixture of conditionally autoregressive (CAR) priors, which are frequently used in time series, spatial statistics and image processing (e.g., see \citealp{rue2005gaussian}).  By assuming that $\theta_{i,j,0,r}$ is known we ensure that the prior distribution, and therefore the associated posterior, is proper.  Indeed, note that the more common choice of a flat (improper) prior on $\theta_{i,j,t,r}$ leads in this case to an improper posterior distribution \citep{sun1999posterior}. In addition, assuming double exponential priors is equivalent to imposing $L_1$ penalty functions on the differences of the parameters in contiguous time points. This penalty type is commonly referred to as the fused lasso with tuning parameter $\lambda$.  An extensive review of the fused lasso and its theoretical properties is presented in \citet{Rinal09}. 

We propose to set the hyperparameters $\hat{\theta}_{1,0}$, $\hat{\theta}_{2,0}$ and $\hat{\theta}_{3,0}$ using a procedure reminiscent of empirical Bayes. In particular, we assume values of $\hat{\theta}_{1,0}$, $\hat{\theta}_{2,0}$ and $\hat{\theta}_{3,0}$ so that the probabilities of the (unobserved) events $( y_{i,j,0}, y_{j,i,0} ) = ( 0, 0 )$, $( y_{i,j,0}, y_{j,i,0} ) = ( 1, 0 )$, $( y_{i,j,0}, y_{j,i,0} ) = ( 0, 1 )$ and $( y_{i,j,0}, y_{j,i,0} ) = ( 1, 1 )$ correspond to their time-average probabilities, i.e., 
\begin{align*}
\hat{\theta}_{1,0} &= \frac{\hat{p}_{1,0}}{\hat{p}_{0,0}}     &     \hat{\theta}_{2,0} &= \frac{\hat{p}_{0,1}}{\hat{p}_{0,0}}     &     \hat{\theta}_{3,0} &= \frac{\hat{p}_{1,1}}{\hat{p}_{0,0}} - \frac{\hat{p}_{1,0}}{\hat{p}_{0,0}} - \frac{\hat{p}_{0,1}}{\hat{p}_{0,0}}   ,
\end{align*}
where
\begin{align*}
\hat{p}_{0,0} &= \frac{2}{n( n - 1 )}\sum_{t = 1}^{T} \sum_{i = 1}^{I} \sum_{j = i + 1}^{J}\mathsf{I} ( y_{i,j,t} = 0, y_{j,i,t} = 0 )   , \\
\hat{p}_{1,0} &= \frac{2}{n( n - 1 )}\sum_{t = 1}^{T} \sum_{i = 1}^{I} \sum_{j = i + 1}^{J}\mathsf{I} ( y_{i,j,t} = 1, y_{j,i,t} = 0 )   , \\
\hat{p}_{0,1} &= \frac{2}{n( n - 1 )}\sum_{t = 1}^{T} \sum_{i = 1}^{I} \sum_{j = i + 1}^{J}\mathsf{I} ( y_{i,j,t} = 0, y_{j,i,t} = 1 )   , \\
\hat{p}_{1,1} &= \frac{2}{n( n - 1 )}\sum_{t = 1}^{T} \sum_{i = 1}^{I} \sum_{j = i + 1}^{J}\mathsf{I} ( y_{i,j,t} = 1, y_{j,i,t} = 1 )   ,
\end{align*}
and $\mathsf{I}(\cdot)$ represents the indicator function. Other appealing default alternatives are possible, and we use them to study the sensitivity of the model to the prior specification. For example, we could specify $\hat{\theta}_{1,0}$ as the logit of the average probability of an incoming link over the whole history of the network, $\hat{\theta}_{2,0}$ as the logit of the average probability of an outgoing link, and $\hat{\theta}_{3,0} = 0$ to reflect our assumption of no reciprocity a priori. Finally, we also tried setting $\theta_{1,0} = \theta_{2,0} = \theta_{3,0} = 0$, which is consistent with the idea that all categories have the same probability a priori at time 0.

\section{Estimation and Prediction}\label{sec:estimation}

Let $\bfTheta_{i,j}=\{\bfTheta_{i,j,2}, \bfTheta_{i,j,3}, \bfTheta_{i,j,4} \}$ be the vector of all nonzero parameters for the pair of nodes $(i,j)$.
The log-posterior distribution of the parameters is given by
\begin{align}\label{eq:posterior}
\sum_{i < j} \left\{  V_{i,j}(\bfTheta_{i,j}) - \lambda\sum\limits_{r=2}^{4}\|\bfL\bfTheta_{i,j,r}\|_{1} \right\}
\end{align}
where 
\begin{multline*}
V_{i,j}(\bfTheta_{i,j}) = 
\sum_{t=1}^{T} \Big\{ y_{i,j,t} \theta_{i,j,t,1} + y_{j,i,t} \theta_{i,j,t,2} + y_{i,j,t}y_{j,i,t} \theta_{i,j,t,3} \\
- \log \left( 1 + \exp\left\{ \theta_{i,j,t,1}  \right\} + \exp\left\{ \theta_{i,j,t,2}  \right\} + \exp\left\{ \theta_{i,j,t,1} + \theta_{i,j,t,2} + \theta_{i,j,t,3}  \right\} \right)
\Big\} 
\end{multline*}
is the (unpenalized) log-likelihood, $\| \cdot \|_{1}$ denotes the $L_{1}$-norm, and $\bfL$ is a pairwise difference matrix of dimension
$T \times ( T + 1 )$ of the form

\begin{align*}
\bfL=
\begin{bmatrix}
-1 & 1 & 0 & \cdots & 0 & 0 \\
0 & -1 & 1 & \cdots & 0 & 0\\
 \vdots &  \vdots &       \vdots &      \vdots &   \vdots &    \vdots\\
0 & 0& 0 & \cdots &-1 & 1
\end{bmatrix}.
\end{align*}
Given $\lambda$, \eqref{eq:posterior} can be broken down into $n( n - 1 )/2$ estimation problems, each one corresponding to fitting a multinomial regression for each pair of nodes in the network.

In the sequel we focus on algorithms that can be used to solve each of these independent problems, which are then naively implemented in a parallel environment.  First, we describe two different sampling algorithms for full Bayesian inference.  Estimation results with these sampling schemes are identical but we are interested in comparing their efficiency (see section \ref{sec:applications}).  We also present a faster optimization alternative for point estimation and prediction that allows implementation of the model in big data settings.

\subsection{Full Bayesian Inference}\label{sec:Bayes}

In order to perform Bayesian inference with a multinomial likelihood, we exploit the data-augmentation method based on P\'{o}lya-Gamma latent variables proposed by \citet{PolsonScott13}. Using this approach, the multinomial likelihood can be represented as a mixture of normals with P\'{o}lya-Gamma mixing distribution.  This approach allows for a full conjugate hierarchical representation of the model and posterior inference through relatively simple Markov chain Monte Carlo (MCMC) algorithms.

For the Bernoulli case, the contribution of the observation $y_{t} \in \{0,1\}$ to the likelihood can be written as
\begin{align*}
L(\psi_{t})=\dfrac{\exp(y_t \psi_{t})}{1+\exp(\psi_{t})}
 \propto \exp(\kappa_{t}\psi_{t}) \int_{0}^{\infty}\exp\{-\omega_{t}\psi_{t}^{2}/2\}p(\omega_{t})d\omega_{t}
\end{align*}
where $\psi_{t}$ is the log odds of $y_{t} = 1$, $\kappa_{t} = y_{t} - 1/2$ and $p(\omega_{t})$ is the P\'{o}lya-Gamma density with parameters $(1,0)$. Hence, by augmenting the model with the latent variable $\omega_t$, conditional Gaussianity for the Bernoulli likelihood can be easily achieved.  

Similarly, for the multinomial case, conditional on $\omega_{i,j,t,r}$, the full conditional \textit{likelihood} of each $\theta_{t,r}$ is given by
\begin{align*}
L(\theta_{i,j,t,r} \mid \theta_{i,j,t,-r}) \propto \exp \left \{ -\frac{\omega_{i,j,t,r}}{2}(\theta_{t,r} + C_{i,j,t,r})^{2} +\kappa_{i,j,t,r}(\theta_{i,j,t,r}+C_{i,j,t,r}) \right\}
\end{align*}
with 
\begin{align*}
C_{t,1} &= \log \frac{1 + \exp\left\{\theta_{t,2} + \theta_{t,3}\right\}}{1 + \exp\left\{\theta_{t,2}\right\}}       &       \kappa_{t,1}  &= y_{i,j} - 1/2 \\
C_{t,2} &= \log \frac{1 + \exp\left\{\theta_{t,1} + \theta_{t,3}\right\}}{1 + \exp\left\{\theta_{t,1}\right\}}       &       \kappa_{t,2}  &= y_{j,i} - 1/2 \\
C_{t,3} &= \log \frac{\exp\left\{\theta_{t,1} + \theta_{t,1}\right\}}{\exp\left\{\theta_{t,1}\right\} + \exp\left\{\theta_{t,2}\right\}}       &       \kappa_{t,3}  &= y_{i,j}y_{j,i} - 1/2 
\end{align*}
and $\omega_{i,j,t,r} \mid \bfTheta \sim \PG \left(1,\theta_{i,j,t,r} + C_{i,j,t,r} \right)$. In the previous expression, $\PG$ denotes a P\'{o}lya-Gamma distribution. Hence, conditionally on the latent variable $\omega_{i,j,t,r}$ we obtain an augmented Gaussian likelihood with observations  $y^{*}_{i,j,t,r} \sim \normal(\theta_{i,j,t,r},\omega^{-1}_{i,j,t,r})$, where $y^{*}_{i,j,t,r}=\kappa_{i,j,t,r}/\omega_{i,j,t,r} - C_{i,j,t,r}$.  Hereinafter, we simplify notation by dropping the subindex $i$ and $j$ associated with the subject pair.

\subsubsection{Latent Variables Approach}\label{se:latent}

Using the fact that the double exponential distribution can be expressed as a scale mixture of normals with exponential mixing density (\citealp{ParkCas08}) :
\begin{align*}
\frac{a}{2}\exp(-a|x|)=
\int_{0}^{\infty}\frac{1}{\sqrt{2\pi \tau}}\exp\left(\frac{x^2}{2\tau}\right)\frac{a^2}{2}\exp\left(-\frac{a^{2}\tau}{2}\right)d\tau, 
\end{align*}
the proposed model can be expressed as a simple hierarchical extension of a dynamic linear model 
\begin{align*}
 y^{*}_{t,r}&=\theta_{t,r}+\epsilon_{t,r},  &\epsilon_{t,r}&\sim \normal(0,\omega^{-1}_{t,r}),\\
\theta_{t,r}&=\theta_{t-1,r}+\varepsilon_{t,r}, & \varepsilon_{t,r}&\sim \normal(0,\tau^{2}_{t,r}),
\end{align*}
for $2 \leq t \leq T$, where $\theta_{1,r} \sim \normal\left( \hat{\theta}_{0,r}, \tau^2_{1,r} \right)$ and $\tau^2_{t,r}$ is exponentially distributed a priori with mean $\frac{2}{\lambda^2}$.

We rely on the dynamic linear model representation to update the parameters in a component-wise fashion using a forward filtering backward sampling (FFBS) algorithm (\citealp{Sylvia94,CarterKohn94}).  Furthermore, the latent parameters $\tau_{t,r}$ for $t = 0,\ldots,T-1$ are independent a posteriori and updated as 
\begin{align*}
\left(1/\tau^{2}_{t,r}| \bfTheta_{r},\lambda\right) \sim \IG\left(\sqrt{\frac{\lambda^{2}}{(\theta_{t,r}-\theta_{t-1,r})^{2}}},\lambda^{2}\right), 
\end{align*}
where $\IG$ denotes the Inverse Gaussian distribution (\citealp{KyGiCa10}).

\subsubsection{Direct Sampling}\label{se:direct}

Note that the full conditional prior on $\theta_{t,r}$ only involves its two nearest neighbors,
so that for  $1 \leq t \leq T-1$:
\begin{align*}
\pi(\theta_{t,r}|\theta_{t-1,r},\theta_{t+1,r}) \propto
 \exp \left \{-\lambda( |\theta_{t,r}-\theta_{t-1,r}|+|\theta_{t+1,r}-\theta_{t,r}|)\right\}.
\end{align*}
Hence, the full conditional posterior distribution of $\theta_{t,r}$ is a mixture of truncated normal distributions with three components:
\begin{multline*}
(\theta_{t,r} \mid y^{*}_{t,r},\theta_{t-1,r},\theta_{t+1,r},\omega_{t,r}) \sim
w_{1}\TN(\mu^{(1)}_{t,r},\sigma_{t,r};\theta_{t,r}<\xi_{t,r}) + 
w_{2}\TN(\mu^{(2)}_{t,r},\sigma_{t,r};\theta_{t,r}>\zeta_{t,r}) \\
+ w_{3}\TN(\mu^{(3)}_{t,r},\sigma_{t,r};\xi_{t,r}<\theta_{t,r}<\zeta_{t,r})    
\end{multline*}
where $\sigma_{t,r}=1/\sqrt{\omega_{t,r}}$, $\xi_{t,r}=\min\{\theta_{t-1,r},\theta_{t+1,r}\}$, $\zeta_{t,r}=\max\{\theta_{t-1,r},\theta_{t+1,r}\}$, 
the means of the truncated normal distributions are given by the following expressions:
\begin{align*}
\mu^{(1)}_{t,r} &= y^{*}_{t,r}+\frac{2\lambda}{\omega_{t,r}},      &      \mu^{(2)}_{t,r} &= y^{*}_{t,r}-\frac{2\lambda}{\omega_{t,r}},      &      \mu^{(3)}_{t,r} &= y^{*}_{t,r},
\end{align*}
and the conditional posterior probabilities of the components of the mixture are given by  
\begin{gather*}
w_{1}=\exp\left\{\frac{\omega_{t,r}}{2}\mu^{(1)}_{t,r}-\lambda(\xi_{t,r}+\zeta_{t,r})\right\} \Phi\left(\frac{\xi_{t,r}-\mu^{(1)}_{t,r}}{\sigma_{t,r}}\right)\\
w_{2}=\exp\left\{\frac{\omega_{t,r}}{2}\mu^{(2)}_{t,r}+\lambda(\xi_{t,r}+\zeta_{t,r})\right\}\Phi\left(\frac{-\zeta_{t,r}-\mu^{(2)}_{t,r}}{\sigma_{t,r}}\right)\\
w_{3}=\exp\left\{\frac{\omega_{t,r}}{2}\mu^{(3)}_{t,r}-\lambda(\zeta_{t,r}-\xi_{t,r})\right\}\left[ \Phi\left(\frac{\zeta_{t,r}-\mu^{(3)}_{t,r}}{\sigma_{t,r}}\right)- \Phi\left(\frac{\xi_{t,r}-\mu^{(3)}_{t,r}}{\sigma_{t,r}}\right)\right]   , 
\end{gather*}
where $\Phi$ represents the Gaussian cumulative distribution function.

The direct characterization of the posterior distribution for our model is similar to the work of \citet{Hans09} on Bayesian lasso regression with Gaussian likelihoods. In principle, the efficiency of this algorithm is limited by the use of the full conditional distributions for posterior sampling. However, this approach avoids the introduction of the latent variables $\{\tau_{t,r}\}$ discussed in section \ref{se:latent}.

\subsubsection{Penalty parameter estimation}\label{se:penaltyMCMC}

The value of the penalty parameter $ \lambda$ has a direct impact on the quality of the estimates and predictions generated by the model. Hence, under the Bayesian version of our model we assign $\lambda$ a Gamma hyperprior, $\lambda \sim \Gam \left( a, b \right)$.  This choice is conditionally conjugate and the full-conditional posterior is simply
$$
\lambda \mid \cdots \sim \Gam \left( a + \frac{ 3 (T-1) n (n-1) }{4} , b + \frac{1}{2}\sum_{t=2}^{T} \sum_{i=1}^{n}\sum_{j=i+1}^{n}\sum_{r=1}^{3}  \left| \theta_{i,j,t,r} - \theta_{i,j,t-1,r} \right| \right)  .
$$

Because the parameters are on the logistic scale we select values of $a$ and $b$ such that, marginally, $\V\left\{ \theta_{i,j,t,r} \mid \theta_{i,j,t-1,r} \right\}$ is no larger than 1 (so that we do not favor link probabilities that are very close to either 0 or 1).  We suggest $a=1$ and $b=1/5$, so that the median of $\V\left\{ \theta_{i,j,t,r} \mid \theta_{i,j,t-1,r} , \lambda \right\}$ is approximately 0.4, and perform a sensitivity analysis to investigate the effect of our choice in the quality of the predictions.

\subsubsection{Link Prediction}

As we discussed in the introduction, one of the goals of our analysis is short-term link prediction. 
For either of our sampling algorithms, Monte Carlo posterior samples of the parameters at a future time $T+1$ can be obtained as:
\begin{align*}
\theta^{(b)}_{i,j,T+1,,r} &\sim \DE \left( \theta^{(b)}_{i,j,T,r},1/\lambda^{(b)} \right)  ,  &    b&=1,\ldots,B.
\end{align*}  
Hence, we can estimate (for $i<j$) the probability of a directed link from node $i$ to node $j$ at time $T+1$ as
\begin{multline*}
\hat{p}\left(y_{i,j,T+1} = 1 \mid \bfY_{T} \right) = 
\dfrac{1}{B}\sum_{b=1}^{B}\left\{p\left[(y_{i,j,T+1},y_{j,i,T+1}) = (1,0) \mid \bftheta^{(b)}_{i,j,T+1}\right] \right. \\
\left. + p\left[(y_{i,j,T+1},y_{j,i,T+1}) = (1,1) \mid \bftheta^{(b)}_{i,j,T+1}\right] \right\},
\end{multline*}  
with a similar expression being valid for $\hat{p}\left(y_{j,i,T+1} = 1 \mid \bfY_{T} \right)$. 

\subsection{Posterior mode estimation}\label{sec:optim}

Markov chain Monte Carlo algorithms allow for full posterior inference on our model, but can be too slow to be of practical applicability in large datasets.  This issue is particularly pronounced in the case of network data because the number of observations grows as the square of the number of nodes.  As an alternative, we develop an optimization algorithm for maximum a posteriori estimation and prediction.  Our algorithm is an extension of the Split Bregman method proposed by \citet{YeXie11} to solve general optimization problems 
with convex loss functions and $L^{1}$ penalized parameters (see also  \citealp{GoldOsher09}). The algorithm is iterative and 
involves the reformulation of \eqref{eq:posterior} as a constrained problem with the linear restriction
$\bfL\bfTheta=\bfb$, and the introduction of a vector of dual variables $\bfv$ used to split the optimization problem into more tractable steps.
Furthermore, we also rely on a second-order Taylor approximation to the multinomial likelihood for the implementation.

The proposed algorithm consists on repeating the following steps until convergence for each
vector of parameters $\bfTheta_{r}$: 
\begin{enumerate} [ (i) ]
\item $ \bfTheta_{r}^{(m+1)}=\underset{\bfTheta}{\arg\max}  \quad V(\bfTheta^{(m)})
-\langle \bfv_{r}^{(m)},\mathbf{L}\bfTheta_{r}^{(m)}-\bfb_{r}^{(m)} \rangle -\frac{\mu}{2}\|\bfL\bfTheta_{r}^{(m)}-\bfb_{r}^{(m)}\|^{2}_2$
\item $\bfb_{r}^{(m+1)}=\mathfrak{T}_{\lambda_{2}\mu^{-1}}\left(\bfL\bfTheta_{r}^{(m+1)}+\mu^{-1}\bfv_{r}^{(m)}\right)$
\item $\bfv_{r}^{(m+1)}=\bfv_{r}^{(m)}+\delta\left(\bfL\bfTheta_{r}^{(m+1)}-\bfb_{r}^{(m+1)}\right)$
\end{enumerate}
where $\bfv_{r}$ is a vector of dual variables, and $\mathfrak{T}_{\lambda}(\mathbf{w})=[t_{\lambda}(w_1),t_{\lambda}(w_2),\ldots]^{'}$
is a thresholding operator with $t_{\lambda}(w_i)=\sgn(w_i)\max\{0,|w_i|-\lambda\}$, and $0<\delta \leq\mu$. We follow previous literature and set $\delta=\mu$ for our implementation noting that convergence of the algorithm is guaranteed for any value of $\mu$ \citep{YeXie11,GoldOsher09}.

Efficiency of this algorithm is mainly constrained by the maximization of ${\bfTheta_r}$ in the first step. To accelerate it, we replace $V(\bfTheta)$ by its second-order Taylor expansion around the current iterate and proceed to perform component-wise optimization (e.g., see \citealp{Krishna05}). Using this substitution, subproblem (i) is differentiable and the estimate of a component $\theta_{t,r}$ of $\bfTheta_{r}$ for  $1 < t < T-1$ is updated as:

\begin{multline*}
\hat{\theta}^{(m+1)}_{t,r}= \left(G^{(m)}_{t,r}-2\mu \right)^{-1} 
\left[G^{(m)}_{t,r}\hat{\theta}^{(m)}_{t,r}-g^{(m)}_{t,r} - (v^{(m)}_{t,r}-v^{(m)}_{t-1,r})\right. \\ \left.
-\mu(\hat{\theta}^{(m)}_{t+1,r}+\hat{\theta}^{(m)}_{t-1,r}+b^{(m)}_{t-1,r}-b^{(m)}_{t,r})\right],
\end{multline*}
where $g^{(m)}_{t,r} = \left. \frac{\partial V}{\partial \theta_{t,r}} \right|_{\bfTheta^{(m)}_{r}}$ and 
$G^{(m)}_{t,r} = \left. - \frac{\partial^2 V}{\partial \theta^{2}_{t,r}} \right|_{\bfTheta^{(m)}_{r}}$ are the gradient and the information in the direction of $\theta_{t,r}$ evaluated in the current iterate values. The updates for $t=T$ are obtained in a similar fashion with some minor adjustments.

Note that in the maximum a posteriori estimates obtained in this fashion, the coefficient differences ($\bfb=\bfL\bfTheta$) can be exactly zero. This induces a block partition of the parameters that is suitable for change-point identification \citep{RojasWahl14, HarLevy08}.   

\subsubsection{Selection of the penalty parameter}\label{se:select}

The penalty $\lambda$ can be selected through cross-validation by training the model on an observed sample $\bfY_1, \ldots, \bfY_t$, and performing a one-step-ahead prediction for $\bfY_{t+1}$ for a grid of values of $\lambda$. This procedure can be repeated to obtain a set of predicted networks $\hat{\bfY}_{t+1}, \ldots, \hat{\bfY}_{t+m}$ for $t+m \leq T$, each of these predictions can then be compared against the respective observed networks, the number of false and true positives is computed, and a receiver operating characteristic (ROC) curve is constructed. Finally, the optimal penalty parameter can be chosen as the value of $\lambda$ in the grid that provides the highest area under the curve (AUC) average over the $m$ predicted networks in the testing dataset.   

One potential drawback of this approach is that selection of the optimal tuning parameter through cross-validation can be computationally expensive \citep{Tibshi05}.   A popular alternative method that can be used with our MAP estimation procedure is to use model selection criteria (e.g AIC, BIC).  Our approach is to select the penalty $\lambda$ from among a pre-specified grid of values by maximizing the Bayesian Information Criteria (BIC)
\begin{align*}
BIC_{\lambda} =\sum_{i < j} \left[ 2V_{i,j}(\hat{\bfTheta}_{i,j}) - \mathcal{K}_{i,j}(\lambda) \log(T-1)\right],
\end{align*}
where $\mathcal{K}_{i,j}(\lambda)$ is an estimate of the number of degrees of freedom when the penalty parameter $\lambda$ is used to compute the MAP estimate.  In the case of the fused lasso, \citet{TibTay11} showed that the number of non-zero blocks of coefficients in $\hat{\bfTheta}_{i,j}$ is a rough unbiased estimate of the degrees of freedom.

\subsubsection{Link Prediction}

Given a point estimate $ \hat{\bftheta}_{i,j,T}$ based on an observed sample $\bfY_1, \ldots, \bfY_T$,
the probability of a directed link from node $i$ to node $j$ at time $T+1$ is estimated as
\begin{multline*}
\hat{p}\left(y_{i,j,T+1} = 1 \mid \bfY_{T} \right) =
p\left[(y_{i,j,T+1},y_{j,i,T+1}) = (1,0) \mid \hat{\bftheta}_{i,j,T}\right]\\
+ p\left[(y_{i,j,T+1},y_{j,i,T+1}) = (1,1) \mid \hat{\bftheta}_{i,j,T}\right],
\end{multline*}  
with a similar expression being valid for $\hat{p}\left(y_{j,i,T+1} = 1 \mid \bfY_{T} \right)$. 

\section{Related Work}\label{sec:related}

\subsection{Computation}\label{se:revcomput}

The literature on algorithms for parameter estimation for linear regression with a fused lasso penalty is extensive. This is a challenging problem because the fused lasso penalty is not a separable and smooth function, and traditional optimization methods fail under these conditions.  In particular, some algorithms that provide a solution path for sequential increments of the regularization parameter have been developed for the Fused Lasso Signal Approximator (FLSA)  where the design matrix is $\bfX=\bfI$  \citep{Fried07, Hoef10a}, and for a general full rank design matrix  $\bfX$ \citep{TibTay11} only in the case of gaussian regression. 

In this work, we are interested in fused lasso penalized multiclass logistic regression.   \citet{FriHaTib10} explores coordinate descent regularization paths for logistic and multinomial logistic regression by using iteratively reweighted least squares (IRLS) but only for lasso, ridge and elastic net penalties (see also \citealp{Krishna05}). \citet{Hoef10b} proposes a coordinate-wise algorithm for the fused lasso that can be extended to logistic regression using iterative reweighted least squares (IRWLS), but no path solution algorithms have been fully developed for the multinomial logistic regression setting that is the focus of this paper. Recently, \citet{YuWon13} introduced a Majorization-Minimization (MM) algorithm for fused lasso penalized generalized linear models that benefits from parallel processing. They also present a good comparison with other existing algorithms including regularization path and first-order methods. For a fixed set of penalization parameters, several optimization algorithms have been proposed for fused lasso problems with general smooth and convex loss functions but not for the specific case of multinomial logistic regression. \citet{LiuYuYe10} proposes an Efficient Fused Lasso Algorithm (EFLA) which solves a FLSA subproblem via a Subgradient Finding Algorithm. \citet{GoldOsher09} use the split Bregman iteration method to deal with a set of image processing problems that can be treated as general $L_1$ penalized problems. Motivated by this idea, \citet{YeXie11} developed the split Bregman based algorithm for the generalized fused lasso with Gaussian likelihoods. We further extend split Bregman algorithms by introducing a version of the approach for categorical and, in particular, dyadic data likelihoods.  In our experience, these kind of algorithms tend to converge faster and avoid local modes that offer difficulties to most of the other algorithms mentioned above.

From a Bayesian perspective, a general hierarchical model for penalized linear regression that includes the fused lasso penalty is presented in \citet{KyGiCa10} for the Gaussian case (see also \citealp{ParkCas08, Hans09}). In contrast, the MCMC algorithms discussed in Section \ref{sec:Bayes} are designed to deal with categorical data.  Furthermore, the latent variable approach from Section \ref{se:latent} exploits the particular Markovian structure of the problem at hand to generate a much more efficient algorithm than the naive implementation of \citet{KyGiCa10} would suggest.  On the other hand, and to the best of our knowledge, the direct sampling algorithm of Section \ref{se:direct}, which extend that of \cite{Hans09} from the regular to the fussed lasso has never been described in the literature before. It is also worth mentioning the work of \citet{ScottPillow12}, who used a data augmentation approach for full Bayesian inference of neural spike data counts observed over time by proposing a dynamic negative-binomial factor model with an autoregressive structure. Although both kinds of problems share time-dependent parameters and their algorithm shares some features with our latent variable sample, the structure of our dyad-based likelihood is quite different, and the P\'{o}lya-Gamma augmentation scheme required for our network represents a non-trivial extension.

Alternative sparsity inducing priors for time-varying parameters to the fused lasso have been introduced in \cite{chan2012time}, who discuss priors for model selection in dynamic contexts, and by \cite{fruhwirth2010stochastic}, \cite{kalli2014time} and \cite{belmonte2014hierarchical}, who derive a continuous shrinkage prior that aggressively shrink small coefficients without explicitly zeroing them out.  All these techniques were developed in the context of dynamic regression models.  Although they could be adapted to identify change points by considering differences between parameter levels, implementing them would come at a significant additional computational cost.

\subsection{Models for dynamic networks}

\citet{Sarkar12} presents a nonparametric link prediction algorithm for sequences of directed binary networks where each observation  in time is modeled using a moving window, and the function is estimated through kernel regression. They also incorporate pair specific features, and a spatial dimension using local neighborhoods for each node.  \citet{HuangLin09} present an autoregressive integrated moving average model (ARIMA), and combine it with link occurrence scores based on similarity indices of network topology measures for link prediction in temporal weighted networks (see also \citealp{daSilva12}).   More recently,  \citet{BlissFrank14} proposed a method based on similarity indices and node attributes joined with a covariance matrix adaptation evolution strategy for link prediction in networks with a large number of nodes. 

Other relevant approaches include the dynamic versions of the latent space model of \citet{Hoff2} presented in \citet{SarkarMoore05} and \citet{SewChen15}, and the work of \citet{XiFuSo10} developing the temporal extension of the mixed membership blockmodel first introduced in \citet{Airoldi} for community identification in social networks.  \citet{BetaRodBoyd15} extend the Bayesian infinite-dimensional model of \citet{Kemp} by linking different time periods through a hidden Markov model.  On the other hand, \citet{HanFuXi10} introduces a temporal version of the Exponential random graph model (tERGM) first introduced in \citet{Frank}.  This temporal model can be used to infer links but its prediction ability is poor unless node attributes or dyadic covariates are included in the model in addition to traditional static network statistics (e.g. reciprocity, transitivity and popularity statistics).  \citet{CranDes11} present a more general temporal ERGM that includes node and dyad-level covariates with applications to political science (see also \citealp{SnijSteBunt10}). In this extension, the square root of the indegree and outdegree are added as node attributes at every time point, and functions of past networks can be utilized as a dyadic covariates.

A key feature of our model is its scalability and efficiency.  Because the model structure is relatively simple and dyads are modeled as conditionally independent, estimation and prediction algorithms are fast and can be easily implemented in parallel environments.  This means that our model can more easily be scaled to long series of large networks than those discussed above.  Conditional independence does have some drawbacks.  In particular, although the model directly models reciprocity, it does not explicitly account for transitivity.  In spite of this, the illustrations we present in the following sections suggest that our model is at least competitive and, in some cases, superior from a predictive point of view to other state-of-the-art models.

\section{Illustrations}\label{sec:applications}

The purpose of this section is to evaluate the performance of our model and compare it with the temporal Exponential Random Graph (tERGM) in terms of its link prediction ability.  We used the \texttt{xergm} package in R to estimate the tERGM \citep{LeCranDes14}.  More specifically, the tERGM is estimated with the \textsf{btergm} function, which implements the bootstrapped pseudolikelihood procedure presented in \citet{DesCran12}. The model we fit includes all the typical ERGM terms, the square root of in and out-degrees as node covariates, and the lagged network and the delayed reciprocity to model cross-temporal dependencies.  

We start by evaluating the performance of the two sampling schemes for Bayesian inference using the effective sample size (ESS) and execution time metrics, and comparing their efficiency with the optimization method for posterior mode estimation using simulated data. We then move on to evaluate the predictive capabilities of the models on both simulated and real data examples consisting of $n=71$ actors and $T=201$ observations in time.  For this purpose, we carry out out-of-sample cross-validation exercises where we hold out the last ten weeks in the data set and make one-step-ahead predictions for the structure of the held-out networks.  More specifically, for each $t=191,192,\ldots,200$ we use the information contained in $\bfY_1, \ldots, \bfY_{t}$ to estimate the model parameters and obtain predictions $\hat{\bfY}_{t+1}$.  Using a simple 0/1 utility function, a future link from node $i$ to node $j$ is predicted as $\hat{y}_{i,j,T+1} = \mathbb{I}(\hat{p}\left(y_{i,j,T+1}=1\mid \bfY_{T} \right)>  f)$, for some threshold $f$ that reflects the relative cost associated with false positive and false negative links.  Each of these predictions is compared against the observed network $\bfY_{t+1}$ to construct a receiver operating characteristic (ROC) curve.  For the tERGM, these results are based on 1,000 MCMC simulations with other function parameters left as the default values (see \textsf{btergm} documentation for more details).

\subsection{MCMC Performance}\label{sec:MCMC}

In order to asses and compare the performance of the latent variable FFBS and the direct sampling MCMC algorithms, we simulated data from our model. The parameters across all the pairs of nodes were randomly drawn from double exponential distributions as $\theta_{t,r} \sim \DE(\theta_{t,r-1},1/\lambda)$ with a true concentration parameter value of $\lambda=3$.  As a measure of efficiency, we use the ESS computed as:
\begin{align*}
ESS=\dfrac{B}{1+2\sum_{k=1}^{K}\rho(k)}
\end{align*}
where $B$ is the number of post burn-in samples, $\rho(k)$ is the autocorrelation at lag $k$, and $K$ is the cutoff lag point according to the initial monotone sequence estimator (\citealp{Geyer92}).

We computed the effective sample size and the CPU run time in seconds for each pair of nodes based on 20,000 iterations after a burn-in period of 2,000 iterations. Table \ref{tab:ESS} shows the results obtained by averaging over 5 runs for each sampling scheme, including the relative efficiency of the algorithms standardizing for CPU run time. From these results it is clear the latent parameters scheme for the fused lasso is much more efficient than the direct sampler that uses the mixture of truncated normals. Based on these results, in the following sections we perform time series cross-validation and prediction for the Bayesian approach using the latent variable FFBS algorithm. 

\begin{table}[h]
\caption{Average ESS and CPU times per pair of nodes for MCMC algorithms.} \label{tab:ESS}
\begin{center}
\begin{tabular}{clclc|c}
{\bf Scheme}  &{\bf ESS}&{\bf CPU(s)}& {\bf Rel.ESS} \\
\hline \hline  
Direct       & 350    & 1827.11 & 0.192\\ 
FFBS       &  3171 & 818.77 & 3.878\\
\hline
\end{tabular}
\end{center}
\end{table}
It is also useful to contrast the execution time of the MCMC algorithms with that of the optimization method, which is only 8.03 seconds on average for each pair of nodes using a stopping criteria of $10^{-5}$ for the relative error. Hence, execution times for the MAP algorithm appear to be at least two orders of magnitude smaller than the fastest version of our MCMC algorithms.

\subsection{Simulation studies}

We first evaluate our model using two simulations. In the first setting, the parameters across all the pairs of nodes were randomly drawn from double exponential distributions so that $\theta_{t,r} \sim \DE(\theta_{t-1,r},1/\lambda)$ with a true penalty parameter value of $\lambda=12$ using initial values $\theta_{r,0}=0$. Because the initial value value of  $\theta_{r,0}=0$ implies a relatively high initial link probability and the evolution variance $1/\lambda$ is relatively small, the resulting network is relatively dense (average number of links of 2682 at each time point, out of 4970 possible ties).  A simple descriptive analysis of the networks shows that they also tend to exhibit low reciprocity and high transitivity. 

\begin{figure}
\begin{center}
\includegraphics[width=3in]{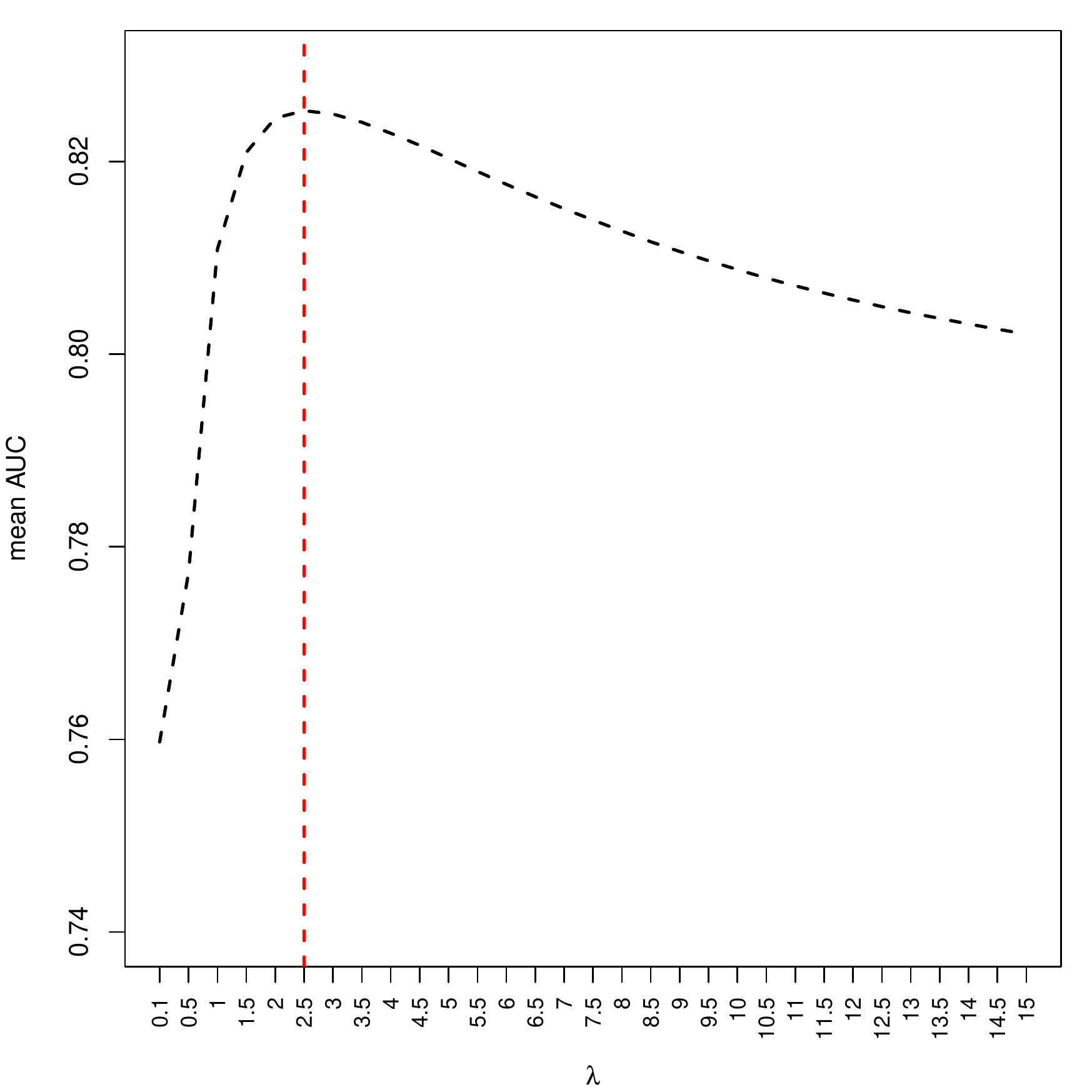}
\includegraphics[width=3in]{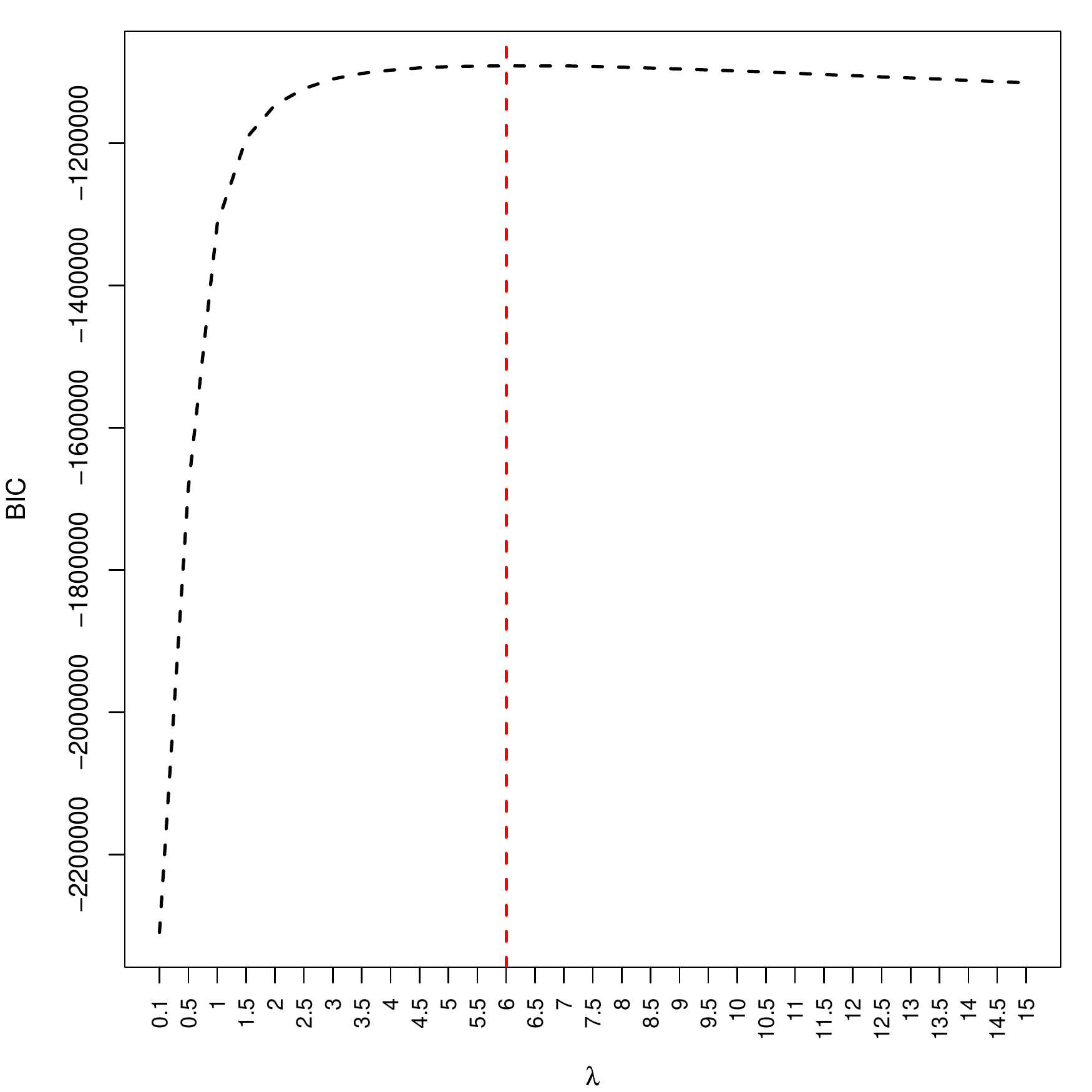}
\end{center}
\vspace{-0.7cm}
\caption{ Simulation 1: Mean AUC over $t=182,183,\ldots,191$ (left panel), and BIC values (right) using the optimization method over a grid of values of $\lambda$ for simulated dataset. The vertical lines indicate the optimal values of $\lambda$.}\label{fig:lambdaSim}
\end{figure}

As discussed in Section \ref{se:select}, we evaluate two methods to select the penalty parameter $\lambda$ for the split Bregman optimization algorithm and evaluate the predictive ability of the model.  Firstly, we use a setup similar to calibration cross-validation (CCV) by partitioning the data into three sets. The first set is used for modeling and consists of the first 181 observations.  Selection of the optimal penalization parameter was performed on the calibration set corresponding to observations $t=182,\ldots,191$, by searching the value of $\lambda$ that maximizes the mean AUC over the predictions of these middle ten observations.  The search for $\lambda$ was conducted over a grid of 31 values between 0.1 and 15; as shown in the left panel of Figure \ref{fig:lambdaSim}, the optimal value is 2.5.  Finally, we report out-of-sample prediction accuracy on the validation set consisting of the last ten observations, $t=192,\ldots, 201$.  Secondly,  we used the first 191 observations to estimate the model and search the value of $\lambda$ that optimizes BIC over the same grid of 31 values between 0.1 and 15. The resulting optimal parameter value in this case is $\lambda=6$ (see right panel of Figure \ref{fig:lambdaSim}).  Again, we evaluate the out-of-sample prediction accuracy of the model on the last ten observations.

\begin{figure}
\begin{center}
\includegraphics[width=3in]{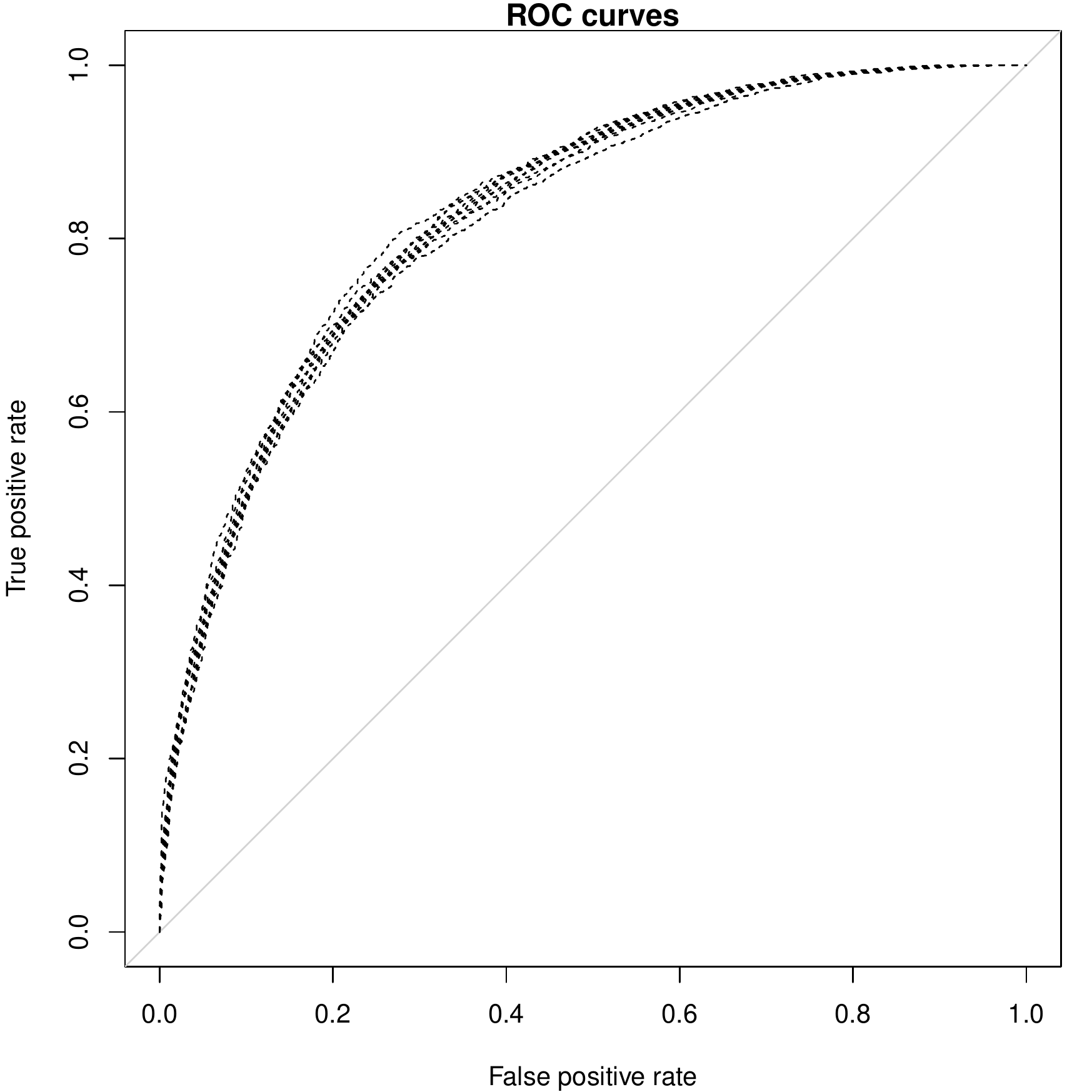}
\includegraphics[width=3in]{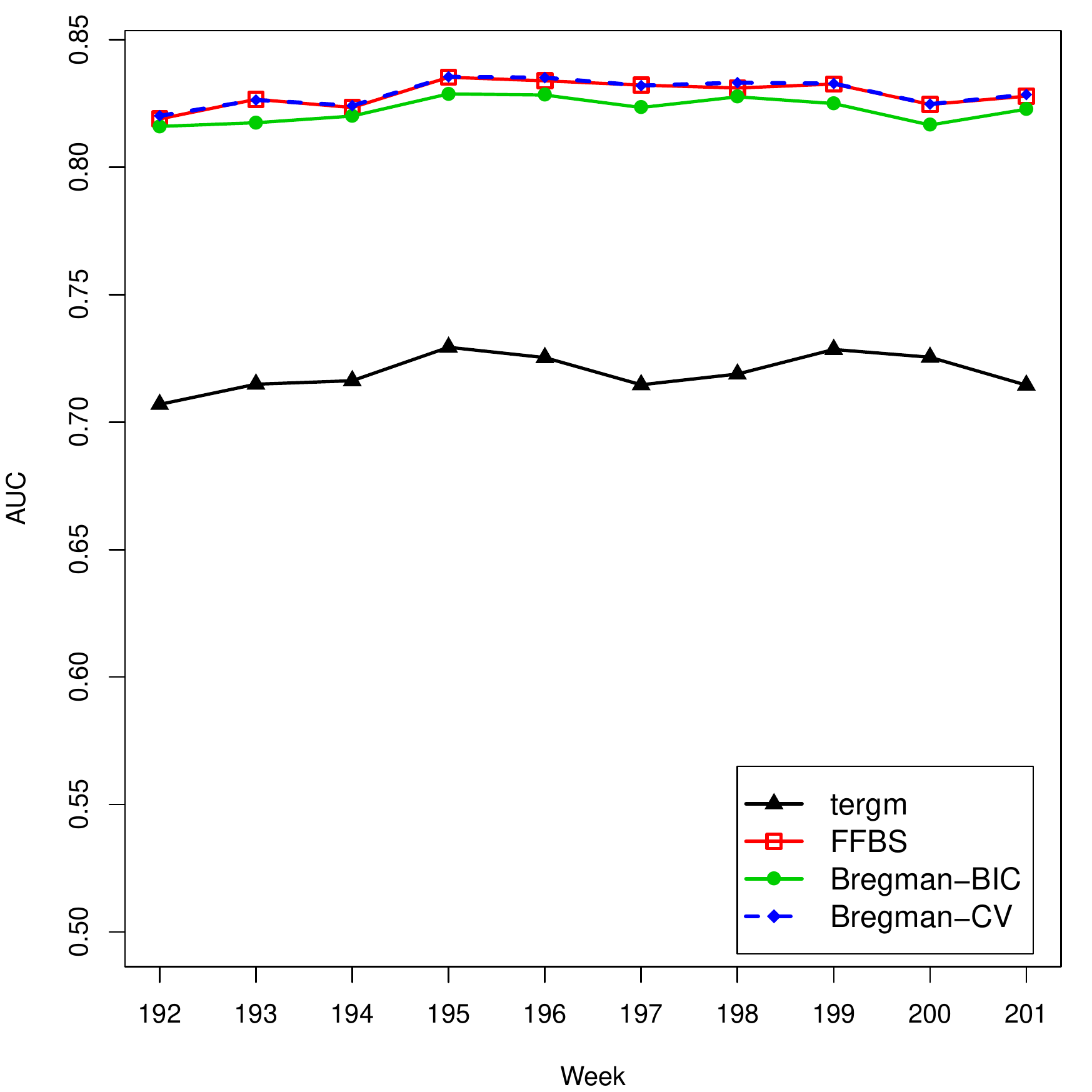}
\end{center}
\vspace{-0.7cm}
\caption{Simulation 1: Plots of the ten operating characteristic curves associated with one-step-ahead out of sample predictions from the fused lasso model with FFBS algorithm (left panel). Area under the curves (AUC) for the temporal ERGM, and the fused lasso model with FFBS algorithm and 
Bregman optimization for simulated data. CV (cross-validation) and BIC represent the two methods for tuning parameter selection.}\label{fig:simula}
\end{figure}

Following Section \ref{se:penaltyMCMC}, for the Bayesian scheme, we used a prior $\lambda \sim \Gam(1, 1/5)$ (mean $5.0$,  95\% prior symmetric credible interval $(0.12, 18)$, which is similar to the range of values used to select $\lambda$ under the optimization algorithm).  The MCMC algorithm is first used to fit our model to the first 191 observations, and then an out-of-sample prediction for observation 192 is generated.  This process is repeated by fitting 192 observations and then predicting observation 193, and so on.  The posterior mean of $\lambda$ is around 9, and varies only very slightly over time.
Figure \ref{fig:simula} shows the ten operating characteristic curves associated with the out-of-sample predictions for the last ten observations using the full Bayesian approach of our model (FFBS algorithm). The right panel of Figure \ref{fig:simula} shows the AUC values for the FFBS approach, the tERGM, and MAP predictions generated by using the optimal value of $\lambda$ obtained from cross-validation (denoted by Bregman-CV) and BIC (denoted by Bregman-BIC). The prediction accuracies for the FFBS algorithm and the Bregman optimization algorithm with cross-validation are almost identical and quite stable over time (both approaches show a good, roughly constant AUC around 83\%). On the other hand, Bregman-BIC performs slightly worse than our two other approaches. Furthermore, in this scenario our model outperforms the tERGM, which shows only a fair predictive performance with an average AUC of 72\%.

\begin{figure}
\begin{center}
\includegraphics[width=3in]{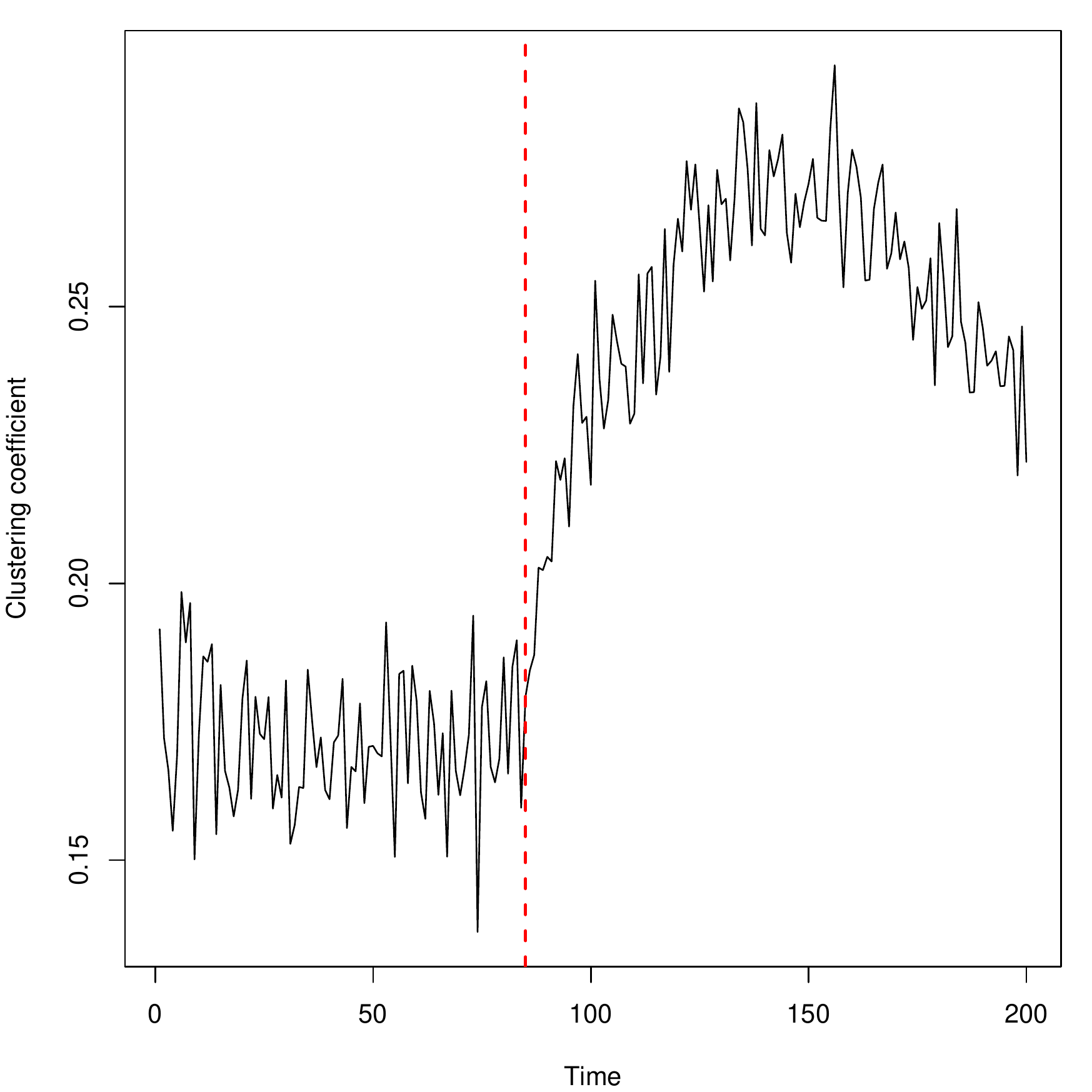}
\includegraphics[width=3in]{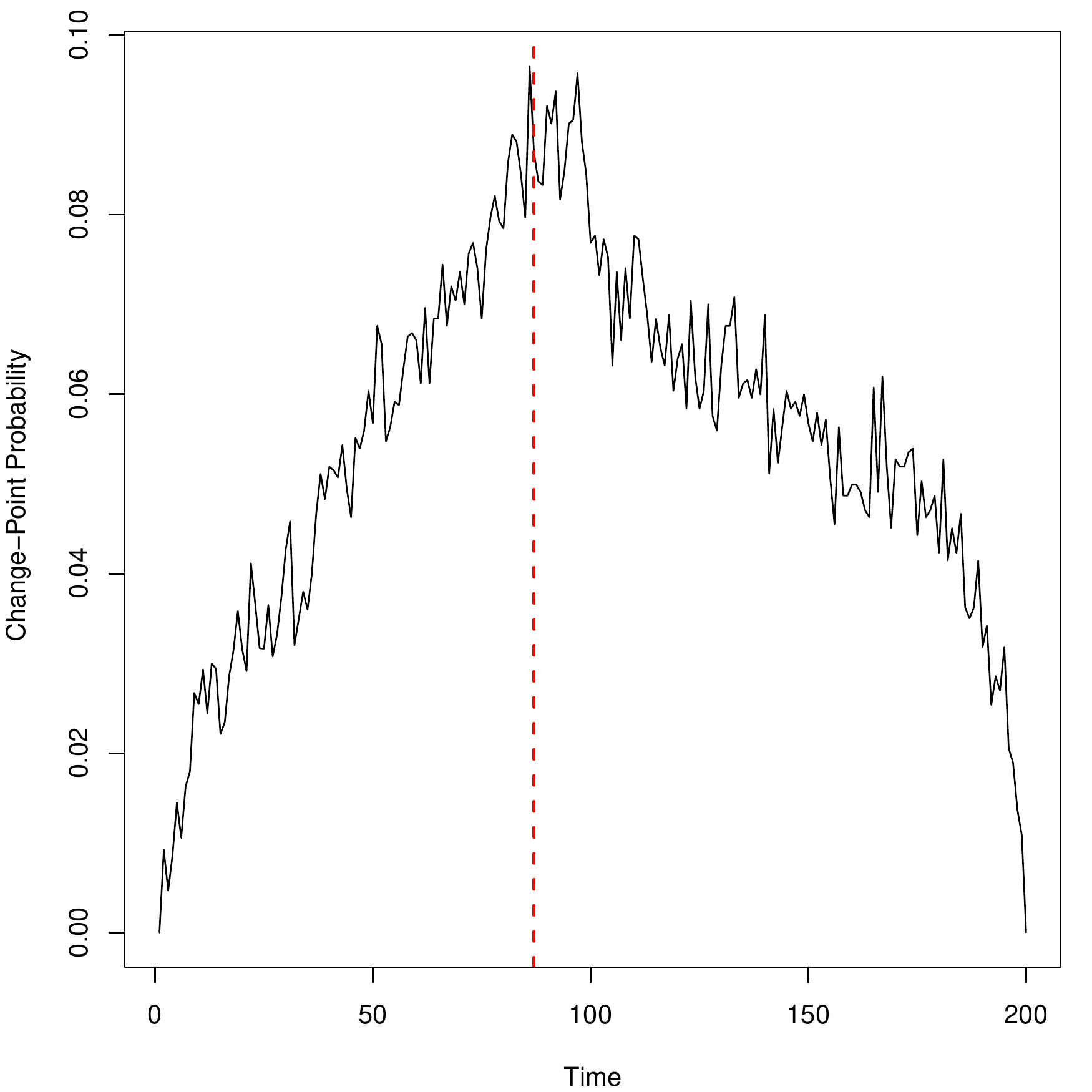}
\end{center}
\vspace{-0.7cm}
\caption{Left panel:  Clustering coefficient for the networks in our second simulated dataset.  Right panel:  Time series of the estimated change-point probability for second simulated data set. The vertical line represents a structural change at time point 85.}\label{fig:simualteddata2}
\end{figure}

For our second simulation, we generated data with similar characteristics to the trading network dataset. The network is sparse with an average of 784 links over time, consistently shows relatively high reciprocity and includes a structural change around time 85, which can be seen in a shift from low to moderate transitivity (see left panel of Figure \ref{fig:simualteddata2}).

\begin{figure}
\begin{center}
\includegraphics[width=3in]{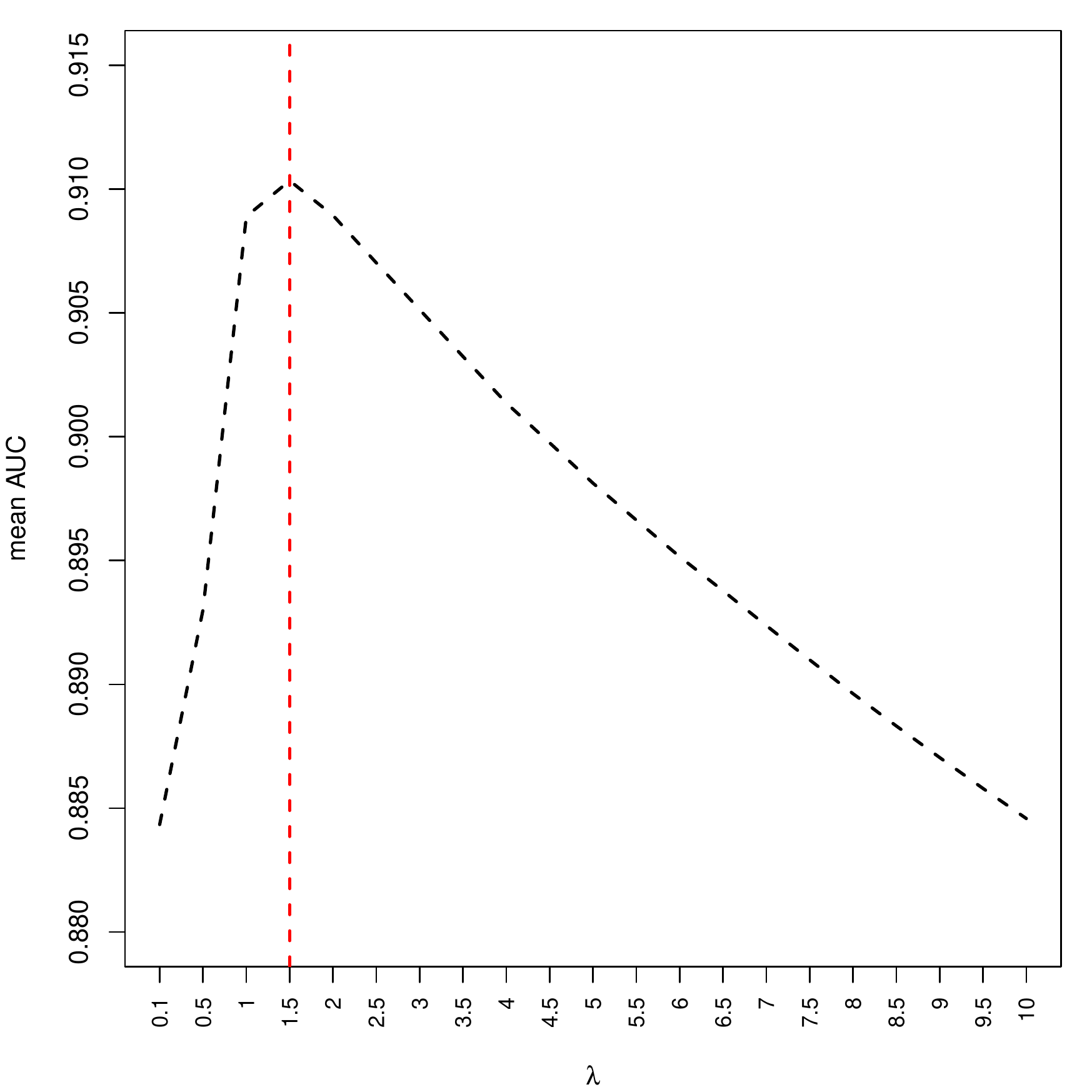}
\includegraphics[width=3in]{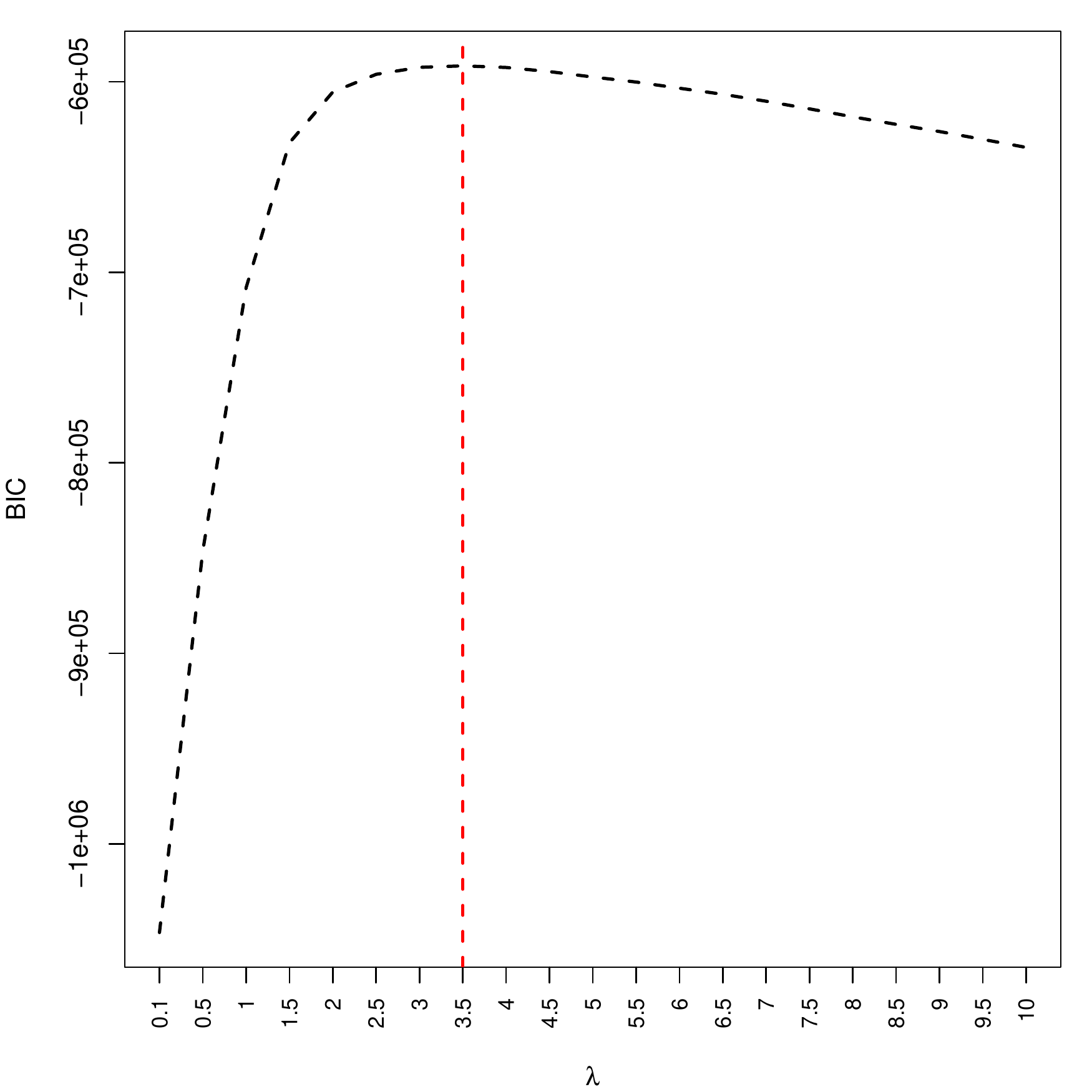}
\end{center}
\vspace{-0.7cm}
\caption{Simulation 2: Mean AUC over $t=182,183,\ldots,191$ (left panel), and BIC values (right) using the optimization method over a grid of values of $\lambda$ for simulated dataset. The vertical lines indicate the optimal values of $\lambda$.}\label{fig:lambdaSim2}
\end{figure}

\begin{figure}
\begin{center}
\includegraphics[width=3in]{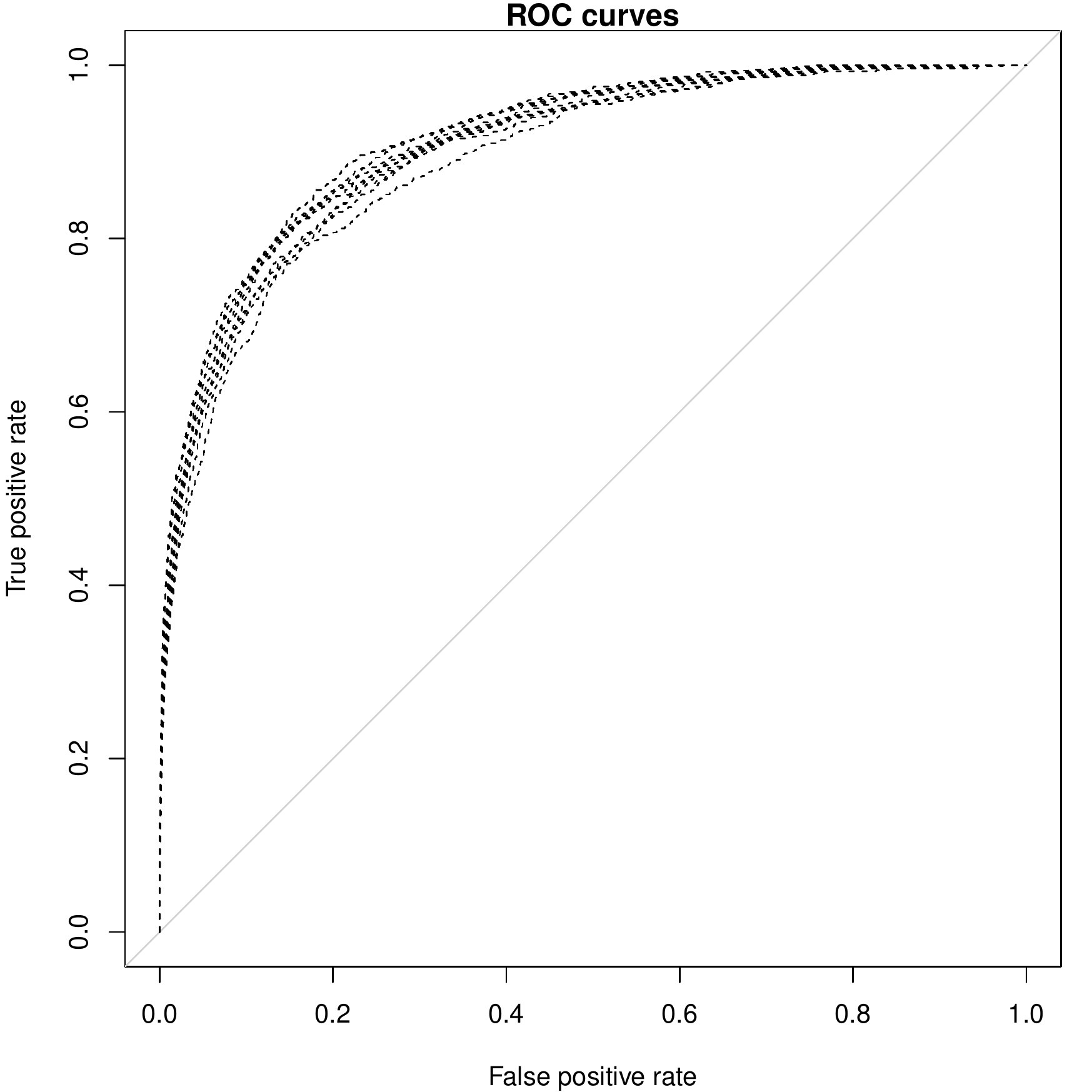}
\includegraphics[width=3in]{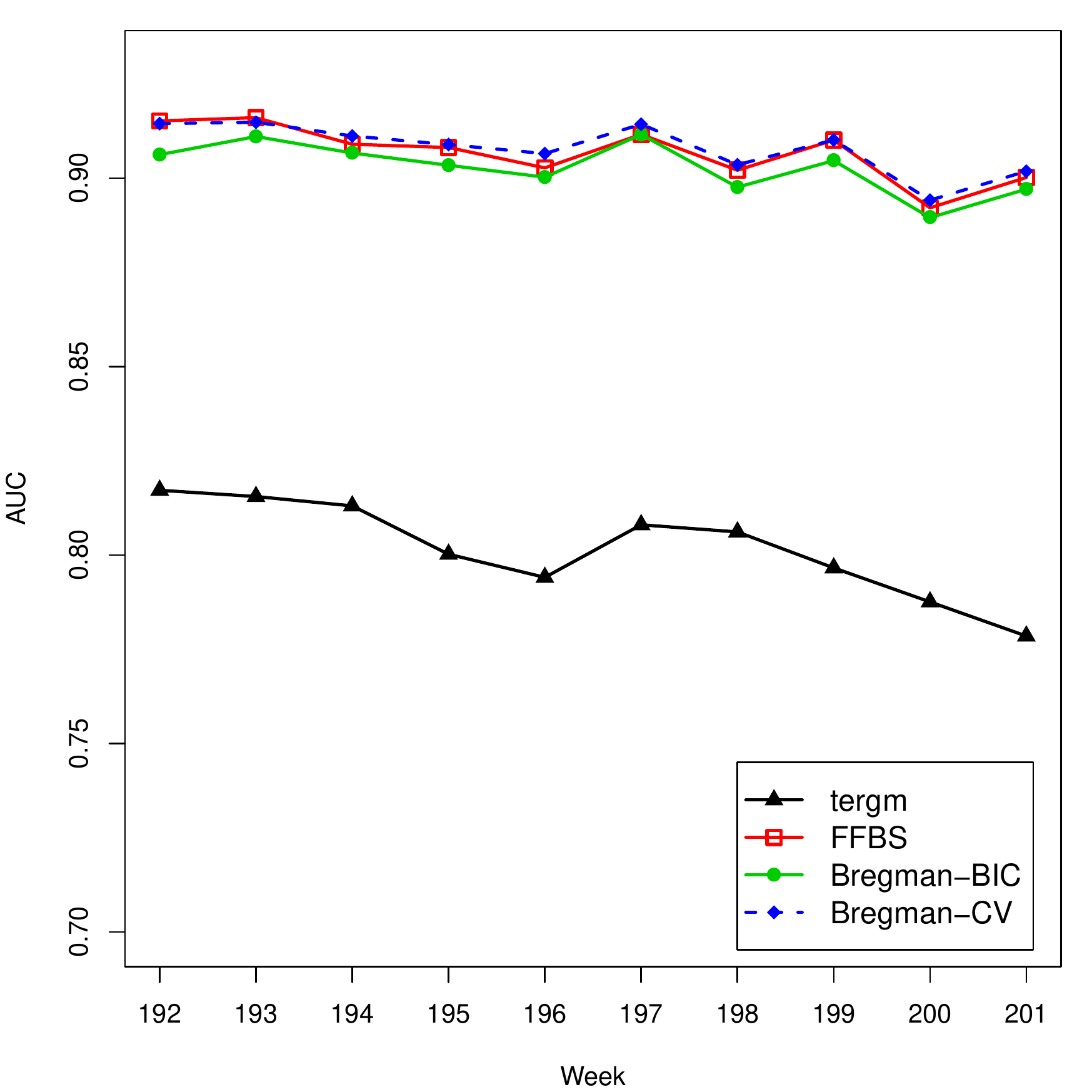}
\end{center}
\vspace{-0.7cm}
\caption{Simulation 2: Plots of the ten operating characteristic curves associated with one-step-ahead out of sample predictions from the fused lasso model with FFBS algorithm (left panel). Area under the curves (AUC) for the temporal ERGM, and the fused lasso model with FFBS algorithm and 
Bregman optimization for simulated data. CV (cross-validation) and BIC represent the two methods for tuning parameter selection.}\label{fig:simula2}
\end{figure}

In this case, the search for the optimal penalization parameter for the optimization algorithm was performed by searching the value of $\lambda$ over a grid of 21 values between 0.1 and 10. Figure \ref{fig:lambdaSim2} shows that the optimal values using the optimization algorithm are $\lambda=1.5$ for cross-validation, and $\lambda=3.5$ using BIC. For the Bayesian approach, assuming a hyperprior $\lambda \sim \Gam(1, 1/5)$, the  posterior mean for $\lambda$ over all pairs of nodes is $3.7$.
Figure \ref{fig:simula2} shows the ten operating characteristic curves associated with the out-of-sample predictions for the last ten observations using the full Bayesian approach of our model (FFBS algorithm). The right panel of Figure \ref{fig:simula2} shows the AUC values for the tERGM and the different algorithms for our model. As before, the prediction accuracies for the FFBS algorithm and the Bregman optimization algorithm with cross-validation are very good (roughly 91\% for both approaches), and Bregman-BIC performs just slightly worse. In this scenario our model again outperforms the tERGM, which shows a good predictive performance with an average AUC of 80\%.

As we mentioned in section \ref{sec:optim}, the maximum a posteriori estimates of the parameters in the fused lasso regression model can be used to identify changes in the network structure over time. In particular, we use an indicator variable that assigns a value of 1 if at least one of the three parameters for pair $(i,j)$ change from time $t-1$ to time $t$, and 0 otherwise. The fraction of these indicators over all pairs of nodes provides a rough estimate of the chances that a change-point has occurred on a given week $t$.  The right panel of Figure \ref{fig:simualteddata2} shows how that proportion changes over time for our second simulation study, which includes a clear change-point around week 85.  As expected, the proportion of dyads showing changes in their parameters peaks on the week the change-point occurs.

\subsection{Inference for financial trading networks}\label{sec:NYMEX}

In this section we analyze a sequence of $T=201$ weekly financial trading networks constructed from \textit{proprietary} trades in the natural gas futures market on the New York Mercantile Exchange (NYMEX) between January 2005 and December 2008. The directed binary networks were constructed by setting $y_{i,j,t}$ = 1 if there was at least one transaction in which trader $i$ sold a contract to trader $j$ during week $t$. 

One particularity of this market is that futures were traded on the New York Mercantile Exchange (NYMEX) only through traditional open-outcry trades until September 5, 2006, and as a hybrid market that included electronic trading conducted via the CME Globex platform after that date. 
Our analysis focuses on 71 traders we identified as being present in the market (although not necessarily active) during the whole period. These trading network is sparse with an average of 826 links each week, and consistently shows very high reciprocity, moderate transitivity, mixing patterns and community structure \citep{BetaRodBoyd15}. 
\begin{figure}
\begin{center}
\includegraphics[width=3in]{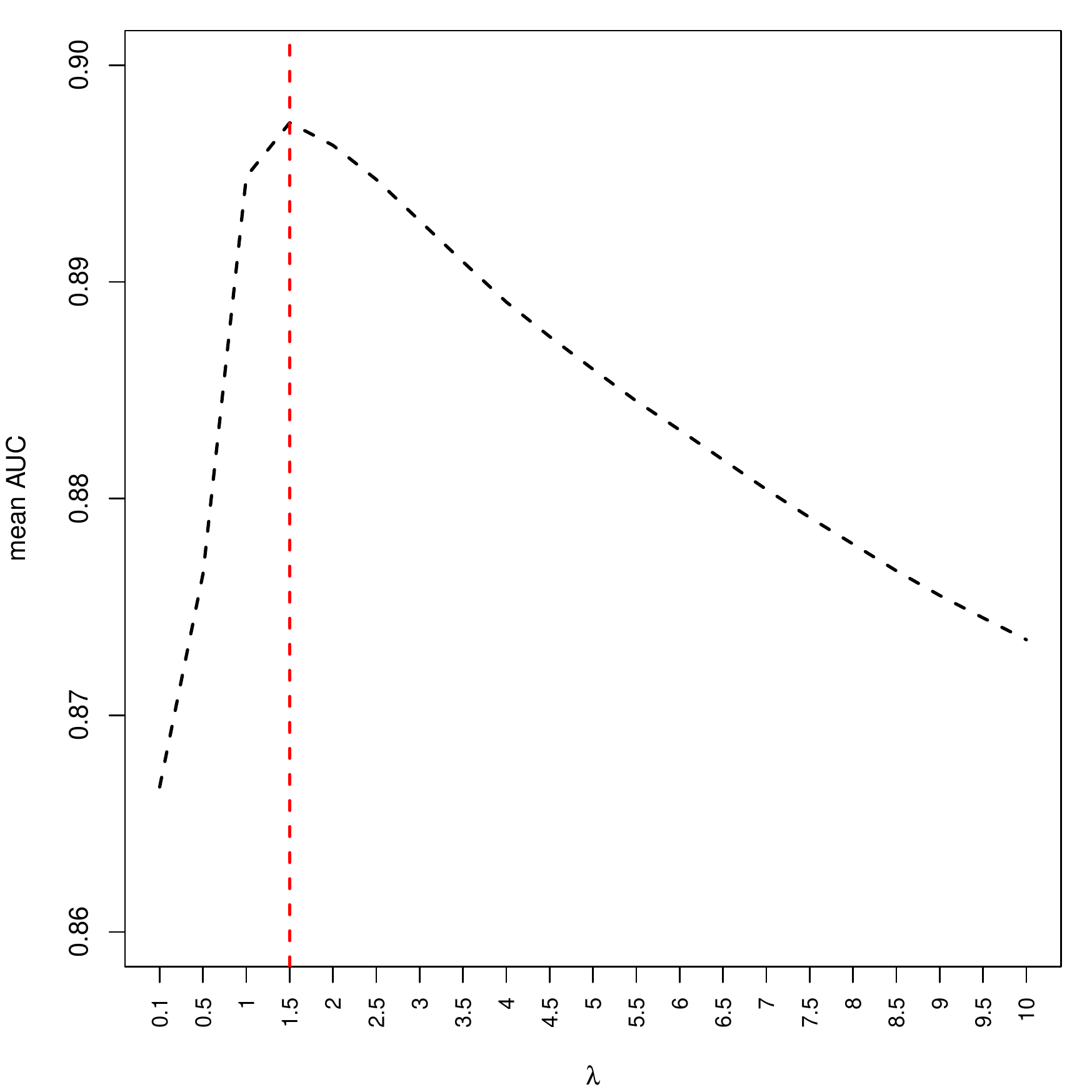}
\includegraphics[width=3in]{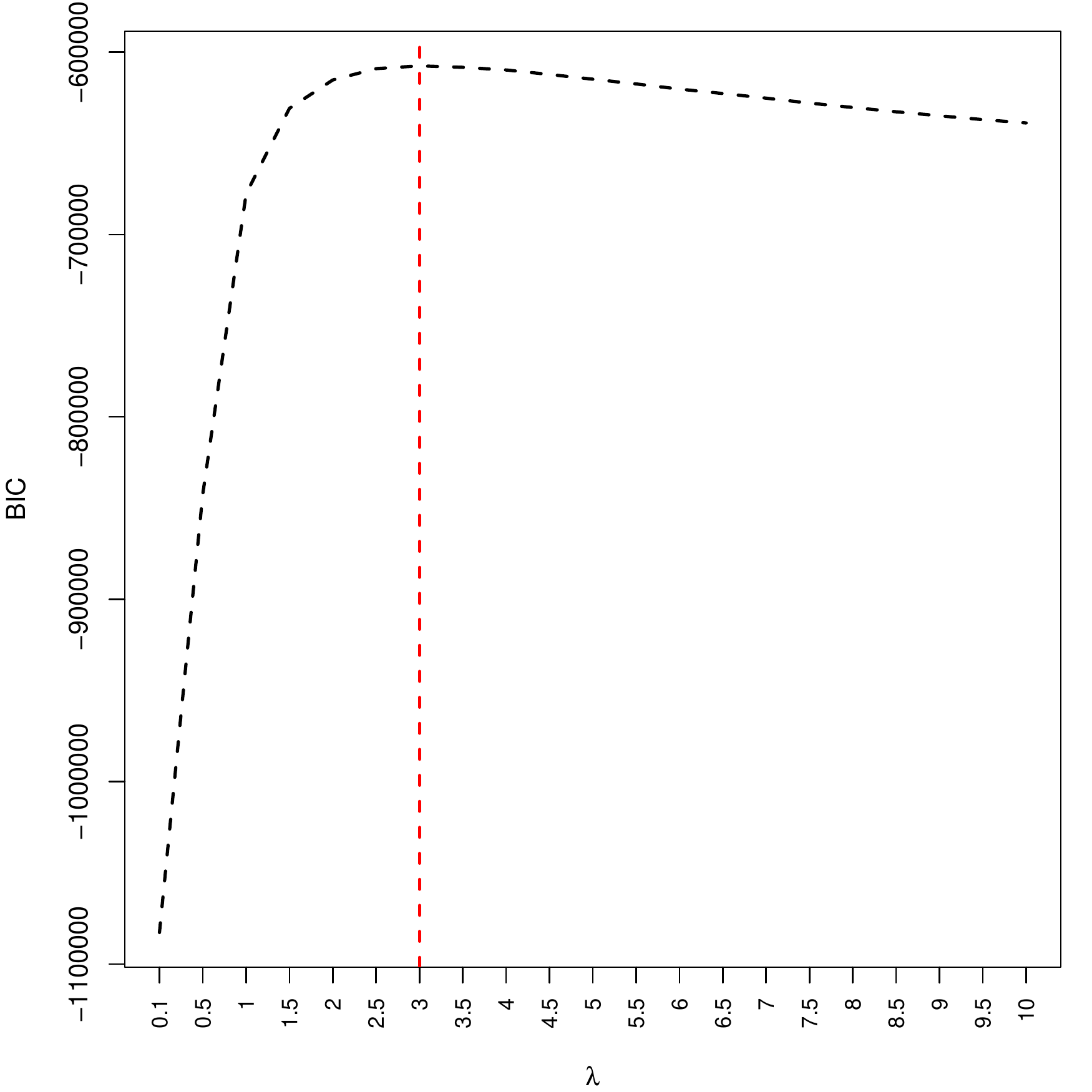}
\end{center}
\vspace{-0.7cm}
\caption{ Mean AUC over $t=182,183,\ldots,191$ (left panel), and BIC values (right) using the optimization method over a grid of values of $\lambda$ for trading network. The vertical lines indicate the optimal values of $\lambda$.}\label{fig:lambda}
\end{figure} 
Analogous to the previous section, selection of the optimal penalization parameter for the optimization algorithm was performed by searching over a grid of 21 values between 0.1 and 10.  The value of $\lambda$ that maximizes the mean AUC over the predictions of ten weeks $t=182,\ldots,191$ (left panel of Figure \ref{fig:lambda}) is $\lambda=1.5$, while the optimal value for BIC over the first 191 weeks is $\lambda=3$. Similarly, for the MCMC algorithm we employ the same $\Gam(1,1/5)$ prior we used in our simulations.
\begin{figure}
\begin{center}
\includegraphics[width=3in]{ROCFFBSNewSimT.pdf}
\includegraphics[width=3in]{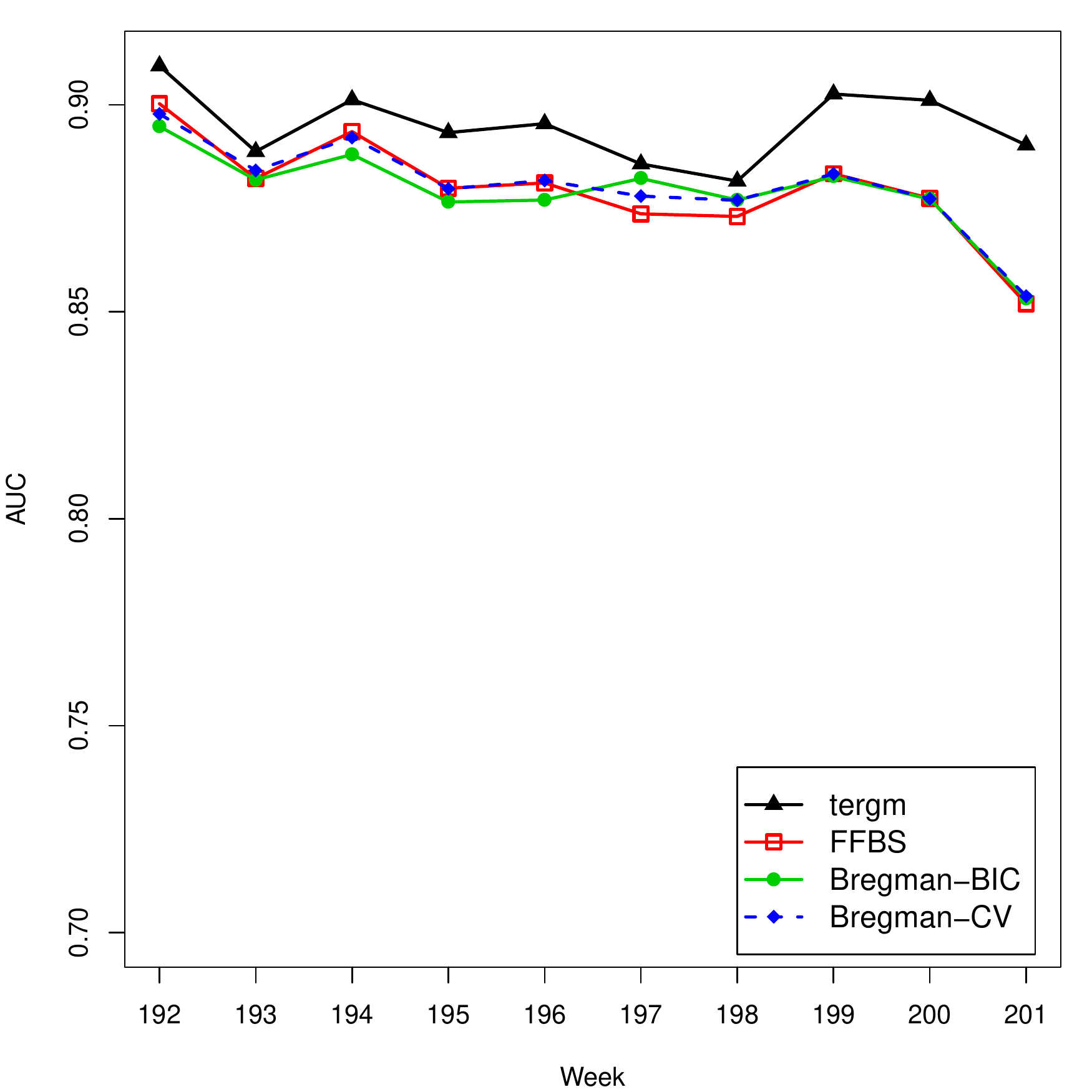}
\end{center}
\vspace{-0.7cm}
\caption{Plots of the ten operating characteristic curves associated with one-step-ahead out of sample predictions from the fused lasso model with FFBS algorithm (left panel). Area under the curves (AUC) for the temporal ERGM, and the fused lasso model with FFBS algorithm and 
Bregman optimization for the trading network. CV (cross-validation) and BIC represent the two methods for tuning parameter selection.}\label{fig:traders}
\end{figure}
The left panel in Figure \ref{fig:traders} shows the operating characteristic curves associated with the out-of-sample predictions generated by our model fitted using the full Bayesian approach. Note that all the curves are very similar, showing that the performance of our model is quite stable over time.  In the same spirit, the right panel of Figure \ref{fig:traders} shows weekly AUCs for the FFBS algorithm, the tERGM, Bregman-CV and Bregman-BIC. The results show that our model performs quite well, with AUC values between 86 to 90\% on every week.  However, in this particular case the tERGM performs slightly but consistently better, with AUC values around 2\% higher.  Furthermore, as in the simulations, the performance of the FFBS and the Bregman-CV algorithm is very similar over all 10 weeks, and the Bregman-BIC performs slightly worse particularly during the first six weeks.

\begin{figure}
\begin{center}
\includegraphics[width=3in]{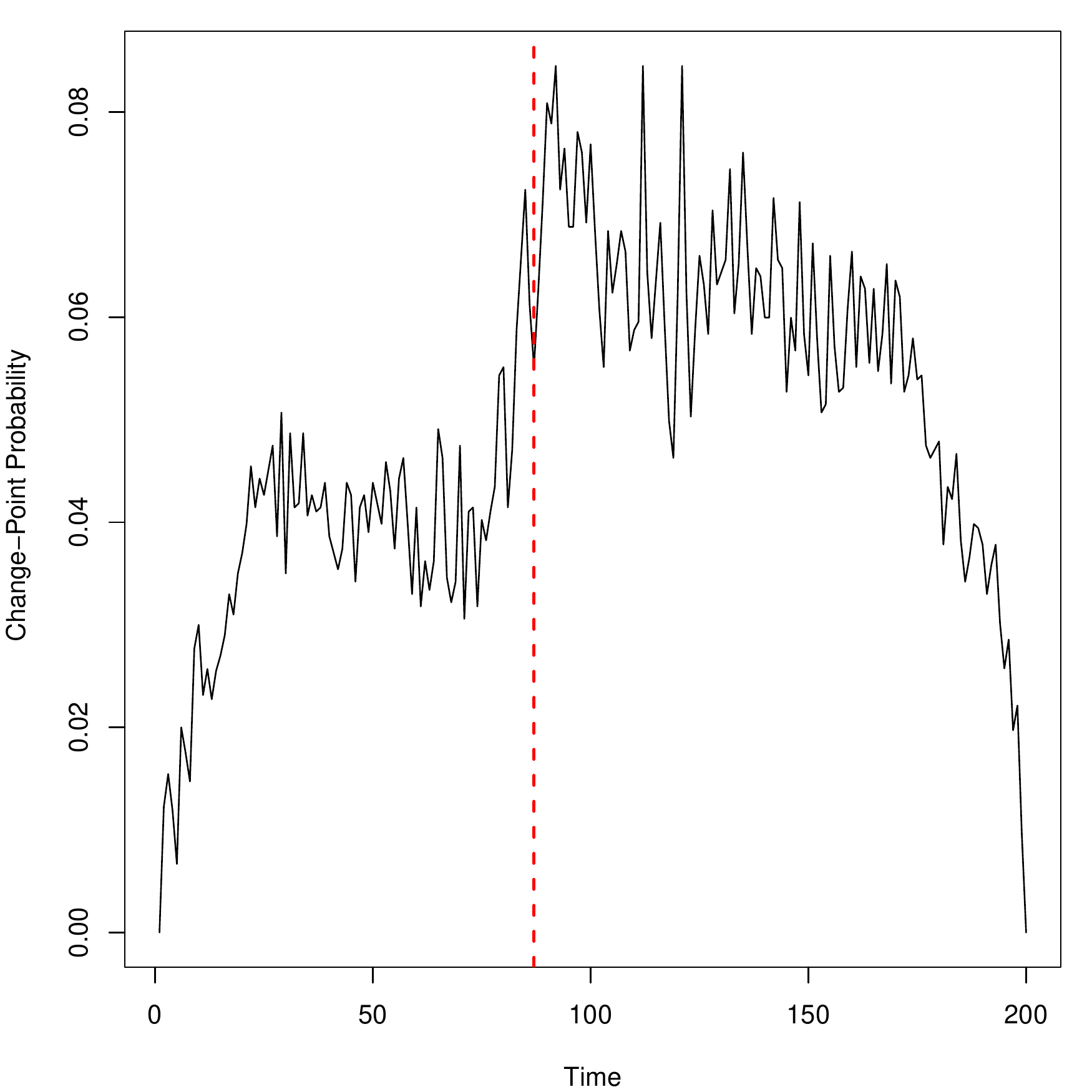}
\end{center}
\vspace{-0.7cm}
\caption{ Time series of the estimated change-point probability
for the trading network. The vertical line represents the introduction of electronic trading in the market at week 85.}\label{fig:change}
\end{figure}

As discussed in our simulation study, the fraction of dyads for which at least one of the three parameters presents a change point at time $t$ provides a rough estimate of the chances that a change-point has occurred on a given week.  Figure \ref{fig:change} presents the time series of the fraction of pairs that show at least one parameter change each week under the Bregman-CV algorithm with the optimal cross-validated $\lambda=1.5$.  The vertical line corresponds to the date of introduction of electronic trading.   Note that the maximum of the time series over the 201 weeks appears right after the introduction of the electronic market and that a second, less marked peak appear around week 124.  These results are consistent with previous analyses of this data \citep{BetaRodBoyd15}.   

\section{Discussion}\label{sec:discussion}

We have discussed a flexible and powerful model for prediction on dynamic networks.  Indeed, the model we present shows competitive performance in the trading network dataset and superior performance in the simulation studies while being much more computationally efficient than alternatives available in the literature.  Furthermore, the model can be easily extended to weighted networks by replacing the multinomial likelihood with an appropriate member of the exponential family.  Similarly, a variation of the model can be devised for undirected networks.

Interestingly, the results on both the simulated and the trading network data showed that the prediction ability of the optimization approach is very similar to that of the Bayesian method, while being far more computationally efficient. In addition, the optimization approach also provides a way of exploring the presence of change points in the network dynamics. On the other hand, the cross-validation approach for tuning parameter selection provides slightly better results than the BIC method, but the computational cost is considerably higher.
  
\bigskip
\begin{center}
{\large\bf SUPPLEMENTAL MATERIALS}
\end{center}

\begin{description}

\item {\bf{\large{\texttt{C++}}} Code}: Code to implement the algorithms for estimation and prediction described in this article. Please refer to the README file contained in the zip file for more details. 
(.cpp files) 
\item {\bf Trading network data set:} Data set used in the illustration in Section \ref{sec:NYMEX}. (.txt file) 

\end{description}

\begin{center}
{\large\bf ACKNOWLEDGMENTS}
\end{center}

This research was partially supported by NSF/DMS award number 1441433.

\bibliographystyle{plainnat}    
\bibliography{Allfinal}

\end{document}